\documentclass[floatfix,aps,twocolumn,showpacs,nofootinbib,superscriptaddress]{revtex4}
\usepackage{graphicx}
\usepackage{epsfig}

\usepackage{amsmath}
\usepackage{amsfonts}
\usepackage{amssymb}
\usepackage{bm}
\usepackage{mathtools}

\usepackage[breaklinks]{hyperref}
\usepackage{balance}
\usepackage{dsfont}
\usepackage{lipsum}
\usepackage{epstopdf}
\usepackage{comment}

\usepackage{pdfpages}
\usepackage{braket}
\usepackage{tikz}
\usetikzlibrary{arrows, shapes}
\usepackage{pgfplots}
\pgfplotsset{compat=1.14}
\usepackage [english]{babel}
\usepackage [autostyle, english = american]{csquotes}
\MakeOuterQuote{"}

\newcommand{\mean}[1]{\left\langle #1 \right\rangle}

\begin{document}

\author{Ashwin Gopal}
\affiliation{Complex Systems and Statistical Mechanics, Department of Physics and Materials Science,
University of Luxembourg, L-1511 Luxembourg, Luxembourg}

\author{Massimiliano Esposito}
\affiliation{Complex Systems and Statistical Mechanics, Department of Physics and Materials Science,
University of Luxembourg, L-1511 Luxembourg, Luxembourg}

\author{Nahuel Freitas}
\affiliation{Complex Systems and Statistical Mechanics, Department of Physics and Materials Science,
University of Luxembourg, L-1511 Luxembourg, Luxembourg}

\title{Large deviations theory for noisy non-linear electronics: \\ CMOS inverter as a case study}

\date{\today}

\begin{abstract}
The latest generation of transistors are nanoscale devices whose performance and reliability are limited by thermal noise in low-power applications. 
Therefore developing efficient methods to compute the voltage and current fluctuations in such non-linear electronic circuits is essential. 
Traditional approaches commonly rely on adding Gaussian white noise to the macroscopic dynamical circuit laws, but do not capture rare fluctuations and lead to thermodynamic inconsistencies. 
A correct and thermodynamically consistent approach can be achieved by describing single-electron transfers as Poisson jump processes accounting for charging effects. 
But such descriptions can be computationally demanding.   
To address this issue, we consider the macroscopic limit which corresponds to scaling up the physical dimensions of the transistor and resulting in an increase of the number of electrons on the conductors. In this limit, the thermal fluctuations satisfy a Large Deviations Principle which we show is also remarkably precise in settings involving only a few tens of electrons, by comparing our results with Gillespie simulations and spectral methods.
Traditional approaches are recovered by resorting to an ad hoc diffusive approximation introducing inconsistencies. 
To illustrate these findings, we consider a low-power CMOS inverter, or NOT gate, which is a basic primitive in electronic design. 
Voltage (resp. current) fluctuations are obtained analytically (semi-analytically) and reveal interesting features.

\end{abstract}

\maketitle

\section{Introduction}


Since the beginning of the information age, the ever-growing need for information processing has fueled spectacular progress in computing technology and the associated semiconductor industry. In a first stage, computing power grew exponentially due to the miniaturization of the MOS transistor, which allowed for higher transistor densities in a chip and higher operation frequencies \cite{ITRS}. However, in the last two decades, the dissipation of power associated with the switching of each transistor made it impossible to continue increasing both the density and the speed. Since then, operation frequencies have stalled, while the number of transistors per chip kept increasing. This enabled the development of multi-core architectures and parallel computing schemes, satisfying in this way the demand for more computing power. This limitation, in addition to the massification of mobile computing devices, has made power consumption the main focus in the development of new computing technologies. A common strategy to reduce power consumption is to reduce the operation voltage, which however also reduces the maximum operation frequency \cite{schrom1996ultra}. In some applications where power consumption is the main concern, and can therefore be traded for speed and performance, the operation voltage can be further reduced. In this way, one enters what is known as the `sub-threshold' regime of operation \cite{wang2006sub}, in which powering voltages can ideally be as low as the thermal voltage $V_{\rm{T}} = k_bT/q_e$, where $q_e$ is the positive electron charge (at room temperature, $V_{\rm{T}} \simeq 26$ mV). However, this trend faces a fundamental limitation, due to the unavoidable presence of thermal noise. A rough estimation shows that the variance in the voltage at the node of a circuit scales as $\sigma^2 \sim k_bT/C$, where $T$ is the temperature at which the circuit works and $C$ the capacitance of the node. The typical values of $C$ are proportional to the scale of the circuit, and in modern fabrication processes they can be as low as $C\simeq 50$ aF, and thus $\sigma \simeq 10$ mV, which cannot be neglected in front of $V_{\rm{T}}$. Therefore, a rigorous description of the intrinsic thermal noise in the electronic circuits used in low-power computing devices is important to study their reliability and performance  \cite{chandra2008impact,freitas2021reliability}. 
More interestingly, understanding how to control intrinsic thermal noise is also crucial in the search of alternative computing schemes where noise is exploited as a resource \cite{kish2009noise,hamilton2014stochastic,camsari2017stochastic, freitas2021stochastic}.

\par
The description of thermal noise in non-linear electronic circuits is commonly done at the phenomenological level, where the intrinsic noise is modelled using a stationary Gaussian white noise, added as an external source \cite{gray2009analysis,nepal2005designing}. Although this description can provide accurate predictions in some applications, it lacks thermodynamic consistency
for non-linear circuits \cite{wyatt1999nonlinear,gupta1982thermal}. This implies that the dynamics could lead to unphysical stationary states, and also may produce violations of the second law of thermodynamics \cite{freitas2021stochastic,brillouin1950can,coram2000thermodynamically}. Moreover, the thermal noise in MOS transistors is of shot noise nature (for sub-threshold operation) \cite{sarpeshkar1993white,landauer1993solid,cui2008measurement}, and the Gaussian approximation fails to capture rare fluctuations in such systems \cite{hanggi1988bistability}. Recently, thermodynamically consistent models have been developed to account for thermal shot noise in nonlinear electronic circuits \cite{freitas2021stochastic,gao2021principles}. In this article, we will be employing the formalism developed in \cite{freitas2021stochastic}. It is based on the idea that conduction through the components of the circuit is due to the transit of excess charges in both forward and reverse directions, and hence it can be modelled using a bi-Poissonian process. The Poisson rates are derived from the I-V curve of the components, constrained by thermodynamic principles, using the modern theory of stochastic thermodynamics. 
Then, this formalism allows to construct a stochastic model of a given electronic circuit by mapping it to a Markov jump process. Similar but thermodynamically inconsistent models have been used before to assess the probability of soft errors in low-power electronic memory caused by thermal noise \cite{li2006model,rezaei2020fundamental}. Common methods to compute the statistics of different dynamical quantities in such models include numerical approaches like the Gillespie simulation of the stochastic dynamics, or spectral methods based on the diagonalization of the generator of the Markov jump process. However, these methods become impractical for large systems and long times. 

\par
\begin{figure*}[ht!]
\centering
\includegraphics[scale=0.7]{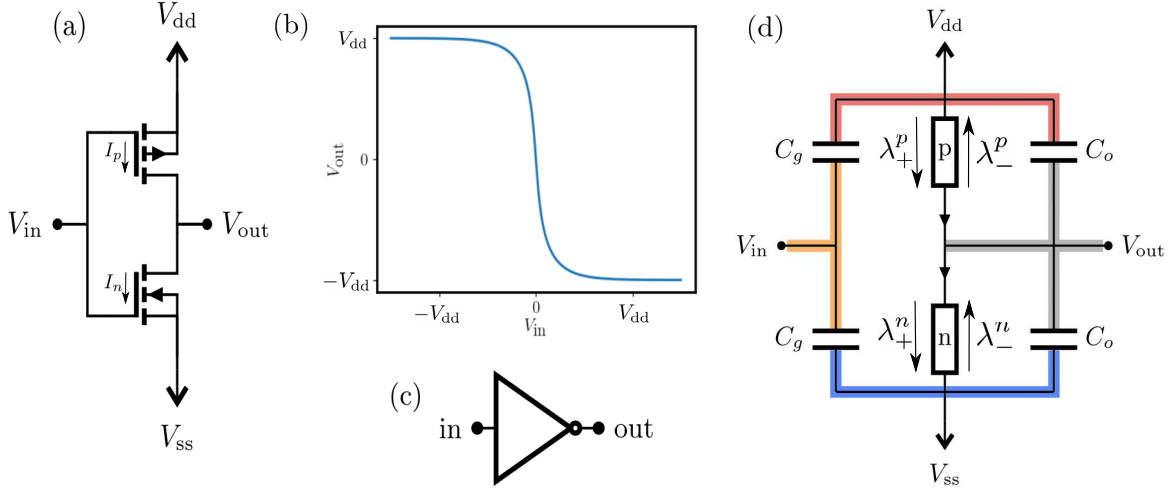}
\caption{(a) The usual implementation of a CMOS inverter. (b) Transfer characteristics of the deterministic subthreshold CMOS inverter, where $V_{\rm{ss}}=-V_{\rm{dd}}$. Parameters : $q_e/q_T=0.1, V_{\rm{dd}}/V_T=3$. (c) Logical symbol for the NOT gate. (d) Jump processes model for the inverter, where the transport of charges are due to the devices (pMOS and nMOS transistors) between different conductors (represented by different colors). }
\label{fig: CMOS inverter}
\end{figure*}

To tackle the above limitations, we will make use of the recent developments in large deviations theory to compute the statistics of observables in the macroscopic regime, by scaling the physical dimensions of the transistors. Large deviations theory deals with the probabilities of rare events, which asymptotically have an exponentially decaying form with a large scale parameter, for e.g. system size, time of observation, number of events, etc. \cite{touchette2009large}. Recently, it has emerged as a formal framework to characterize fluctuations in both equilibrium and out-of-equilibrium systems \cite{derrida1998exact, lebowitz1999gallavotti, giardina2006direct, bertini2015macroscopic, horowitz2017proof}. In equilibrium statistical mechanics, taking the thermodynamic limit allows having a large deviations perspective as a function of system size \cite{ruelle2004thermodynamic,ellis2006entropy}. Whereas for out-of-equilibrium systems, the dynamical aspects are characterized using time-averaged observables, like empirical current, traffic, etc. The long time fluctuations of these observables can be computed using large deviations theory, taking the time of observation as the scale parameter \cite{derrida1998exact,lebowitz1999gallavotti,giardina2006direct}. Another application has been to understand the macroscopic fluctuations of these dynamical observables, using the system size as the scale parameter, analogous to the description in equilibrium systems\cite{bertini2015macroscopic, assaf2017wkb, herpich2020stochastic, freitas2021emergent}. The non-commutativity of the long time and the large size limit is an important feature of systems with dynamical phase transitions, and large deviations theory provides a mathematical framework to characterize them \cite{jack2020ergodicity, vroylandt2020efficiency,lazarescu2019large, meibohm2022finite}.

\par

In this article, we combine the tools of large deviation theory and stochastic thermodynamics to correctly describe the voltage and current fluctuations in electronic circuits due to thermal noise. As a case study, we consider the low-power CMOS inverter (NOT gate), a basic primitive in electronic design. After a thermodynamically consistent description of the inverter, we discuss how the macroscopic limit is naturally approached in these electronic devices, such that the thermal fluctuations satisfy a large deviations principle. Crucially, we show that this limit is reached even in settings involving a few tens of electrons, by comparing the analytical results with the Gillespie simulations and spectral methods.
In this process, we also make comparisons with the widespread diffusive approximations, highlighting its limitations. Finally, we look at the behavior of mean, variance and the precision of the current through the inverter for different parameters, and discuss the conditions under which one can devise a coarse-grained effective model.

 \section{Model for the CMOS inverter }

The CMOS Inverter is an implementation of the most elementary logic gate, the NOT gate. As shown in the Figure~\ref{fig: CMOS inverter}-(a), it is composed of a pMOS (top) and an nMOS (bottom) transistor, with a common gate and drain terminals (with voltages $V_{\rm{in}}$ and $V_{\rm{out}}$, respectively). The circuit is powered by applying a voltage difference between the source terminals of both transistors, $\Delta V= V_{\text{dd}}-V_{\text{ss}}$. In each transistor, electric current flows only between drain and source terminals, due to the presence of a metal-oxide insulator between the gate terminal and the bulk in MOSFETs. To achieve high performance logic designs, these transistors are usually operated in a regime where they essentially behave as a switch. This regime is characterized by a value of input voltage $V_{\text{in}}>V_{\rm{th}}$ (threshold voltage), when there is a proper conduction channel (strong inversion) between source and drain terminals. Whereas, for low-power applications, these transistors are operated in the sub-threshold regime ($|V_{\text{in}}|\le V_{\text{th}}$ ) \cite{wang2006sub}. In this regime, there are still small leakage currents in both the transistors due to partial creation of the conduction channel. These currents have an exponential dependence on the voltages and hence strongly depend on the ratio $V_{\text{in}}/V_{\text{th}}$. We will focus on this mode of operation of the MOS transistors for understanding the role of thermal fluctuations in the CMOS inverter. A CMOS Inverter works as follows. When $V_{\text{in}}<\Delta V/2$, the conduction through the nMOS transistor is reduced, while it is enhanced through the pMOS transistor, and hence the output voltage $v$ quickly relaxes to $V_{\rm{dd}}$. Similarly, for $V_{\text{in}}>\Delta V/2$, the situation is reversed, with the output voltage now quickly relaxing to $V_{\rm{ss}}$ (Figure~\ref{fig: CMOS inverter}(b)). \par
Here, we model the transistors as externally controlled conduction channels between conductors with associated capacitances: $C_g$ (capacitance between the gate electrode and the bulk) and $C_o$ (output capacitance) (See Figure~\ref{fig: CMOS inverter} (d)).  For fixed powering voltages ($V_{\text{ss}}$, $V_{\text{dd}}$) and input gate voltage $(V_{\text{in}})$, the inverter has only one degree of freedom: the voltage on the output conductor (connected to the drain terminals of both transistors), $V_{\rm{out}}\equiv v$. 
\par
In the symmetric powering case, i.e. $V_{\text{ss}}=-V_{\text{dd}}$, the sub-threshold current through the pMOS transistor is given as
\begin{equation}
 I_{\rm{p}}(v;V_{\text{in}},V_{\text{dd}})=I_0e^{(V_{\text{dd}}-V_{\text{in}}-V_{\text{th}})/(nV_{\rm{T}})}(1-e^{-(V_{\text{dd}}-v)/V_{\rm{T}}}),
 \label{eqn: IV}
\end{equation}
where $I_0, V_{\rm{th}}$ and $n$ are the parameters characterizing the transistor.
For such a symmetric case, the current through the nMOS transistor is $I_{\rm{n}}(v;V_{\text{in}},V_{\text{dd}})=I_{\rm{p}}(-v;-V_{\text{in}},V_{\text{dd}}).$ Using Kirchhoff's laws, the deterministic dynamics for $v$ in the CMOS inverter can be obtained as 
\begin{eqnarray}
 2C_o\frac{dv}{dt}=I_{\rm{p}}(v;V_{\text{in}},V_{\text{dd}})-I_{\rm{n}}(v;V_{\text{in}},V_{\text{dd}}).
 \label{eqn: detkitchoff}
\end{eqnarray}
The stationary solution satisfying $dv/dt=0$ gives the transfer characteristics (Figure~\ref{fig: CMOS inverter} (b)), and we will study the deviations from it due to thermal fluctuations.
Since voltages and charges have a linear relation, we can equivalently work in the state space of the output charge $q_{\rm{out}}\equiv q = 2C_o [v +(V_\text{dd} + V_\text{ss})/2]$. The charge transfer between the conductors is modelled as a bi-directional Poisson process. The associated jump rates for p/nMOS transistors are $\lambda_{\pm}^{p/n}(V_{\text{GS}},V_{\text{DS}})$, where, $\pm$ corresponds to the forward ($+$) and reverse ($-$) directions. The elementary voltage change corresponding to a charge transfer $q_e$ is $v_e=q_e/(2C_o)$. The state of the system at any time is characterized by the probability distribution $P(q,t)$ over the charge in the output conductor. The corresponding master equation for the stochastic dynamics of $q$ is given by,
\begin{align}
d_t P(q, t) =&  [\lambda_-^\text{n}(q-q_e) + \lambda_+^\text{p}(q-q_e)]P(q-q_e, t)\label{eq: invertermastereqn}\\
&+  [\lambda_+^\text{n}(q+q_e) + \lambda_-^\text{p}(q+q_e)] P(q+q_e, t)\nonumber\\
&-  [\lambda_-^\text{n}(q) + \lambda_+^\text{n}(q) + \lambda_-^\text{p}(q) + \lambda_+^\text{p}(q)]P(q, t)\nonumber
\end{align}
or
\begin{align}
d_t P(q, t)\! =\! \sum_{\rho}\lambda_\rho(q-\Delta_\rho q_e)P(q-\Delta_\rho q_e,t)\!-\!\lambda_\rho(q)P(q,t),\nonumber
\end{align}
where the summation over $\rho$ in the last line, is over the forward and backward processes($\pm \rho$) of all devices.  \par
The Poisson rates are modelled based on the thermodynamically consistent formalism developed in \cite{freitas2021stochastic}. The rates are obtained from the I-V characteristics of the transistors (Eq.~\eqref{eqn: IV}) along with the thermodynamic constraint of local detailed balance (LDB). This condition on the rates of each device ensures that the system will relax to the equilibrium state at a temperature $T$ corresponding to the environment of that device, if all other devices are disconnected from the circuit. Such a thermodynamic constraint on the dynamics also implies that the fluctuations of different observables are constrained through relations like fluctuations theorems \cite{lebowitz1999gallavotti}  and thermodynamic uncertainty relations \cite{barato2015thermodynamic,horowitz2020thermodynamic}.

For the pMOS transistor of the inverter, the LBD condition reads as
\begin{eqnarray}
 \log\left(\frac{\lambda_+^\text{p}(q)}{\lambda_-^\text{p}(q+q_e)}\right)=-\frac{\delta Q_{q\to q+q_e}}{k_B T},
 \label{eqn: lbdp}
\end{eqnarray}
where $\delta Q_{q\to q+q_e} = U(q+q_e) - U(q) -q_e V_{\rm{dd}}$ is the energy required to perform the transition $q\to q+q_e$. $U(q)= q^2/(4C_0)+ q(V_{\rm{dd}}+V_{\rm{ss}})/2 + \mathrm{const.}$ is the electrostatic energy of the above circuit. Similarly, the LBD condition for the nMOS transistor is given as
\begin{eqnarray}
 \log\left(\frac{\lambda_+^\text{n}(q)}{\lambda_-^\text{n}(q-q_e)}\right)=-\frac{\delta Q_{q\to q-q_e}}{k_B T}.
\label{eqn: lbdn}
\end{eqnarray}
In the subthreshold operation regime (i.e., $V_{\text{in}}<V_{\text{th}}$), the Poisson rates in the inverter for the pMOS transistor after enforcing the LBD condition (Eq.~\eqref{eqn: lbdp}) are: 
\begin{eqnarray}
\lambda_+^\text{p}(q) &=& (I_0/q_e)  e^{(V_\text{dd} - V_\text{in} - V_\text{th})/(\text{n} V_{\rm{T}})}\nonumber\\
\lambda_-^\text{p}(q) &=& \lambda_+^\text{p}(q)  e^{-(-q+q_e/2)/q_T} e^{-\Delta V/(2V_{\rm{T}})}
\label{eqn: rates_p}
\end{eqnarray}
and for the nMOS transistor using Eq.~\eqref{eqn: lbdn} are:
\begin{eqnarray}
\lambda_+^\text{n}(q) &=& (I_0/q_e) e^{(V_\text{in} - V_\text{ss} - V_\text{th})/(\text{n} V_{\rm{T}})}\nonumber\\
\lambda_-^\text{n}(q) &=& \lambda_+^\text{n}(q) e^{-(q+q_e/2)/q_T}e^{-\Delta V/(2V_{\rm{T}})}
\label{eqn: rates_n}
\end{eqnarray}
where we define $q_T=2C_0V_{\rm{T}}.$ The factor $e^{-q_e/(2q_T)}$ in the rates is a consequence of enforcing the LBD condition. Physically, it models the voltage differences due to a single charge transfer (called charging effect), which can become relevant for small devices or at low temperatures \cite{wasshuber2001computational}. The usual modelling using just the I-V characteristics fails to incorporate it \cite{li2006model,rezaei2020fundamental}.\par
In the matrix representation, the master equation Eq.~\eqref{eq: invertermastereqn} can be written as
\begin{equation}
    d_t\ket{P(t)}=\hat{\mathcal{L}}(t)\ket{P(t)}
    \label{eqn: ME_generator}
\end{equation}
where $\ket{P(t)}$ is the probability density vector and $\hat{\mathcal{L}}$ is the generator of the above Markov process, both defined in the orthonormal basis, $\{\ket{q}\}$. The generator, $\hat{\mathcal{L}}$ can be further decomposed into diagonal (escape rates) and non-diagonal parts, such that
\begin{eqnarray}
    \hat{\mathcal{L}} =\hat{\Gamma}(t)-\hat{\gamma}(t),
\end{eqnarray}    
where $\hat{\gamma}(t)$ is a diagonal matrix with escape rates,
\begin{eqnarray}
    \hat{\gamma}(t) &=&\sum_q\ket{q}\bra{q}\left(\sum_\rho\lambda_\rho(q)\right) \hspace{0.15cm}
    \label{eqn: spec_diag}
\end{eqnarray}
and $\hat{\Gamma}(t)$ is a matrix where each off-diagonal term is the jump rate corresponding to that particular transition,
\begin{eqnarray}
    \hat{\Gamma}(t)&=&\sum_q\sum_\rho\ket{q}\bra{q-q_e\Delta_\rho }\lambda_\rho(q-q_e\Delta_\rho ).
    \label{eqn: spec_offdiag}
\end{eqnarray}
Hence, $\hat{\mathcal{L}}(t)$ has a 0 eigenvalue with the left eigenvector $\bra{1}\equiv(1,1,...)$, i.e. $\bra{1}\hat{\mathcal{L}}(t)=0$, which is a property of a Markov generator. The corresponding right eigenvector $\ket{P_{\rm{st}}}$ gives the stationary distribution, $P_{\rm{st}}(q)$, which is equivalent to solving $d_tP(q,t)=0$ from Eqn.~\eqref{eq: invertermastereqn}. The next dominant eigenvalues provide information about the relaxation behavior of the dynamics. Therefore, the spectral decomposition of the generator contains the complete information of the transient and stationary behavior of the stochastic dynamics. We are interested in computing the steady state statistics of observables in such electronic circuits.\par

\par
There are two types of observables in systems described by Markov jump processes. The first kind are state-like observables, which are functions $A(q_t)$ of the state of the system at any given time. Examples of state observables include the voltage of a conductor, internal energy, etc. The stationary fluctuations of those observables can be fully captured by the knowledge of just the stationary distribution $P_{\rm{st}}(q)$. 
The second kind are jump-like observables, which are functionals of both the state
and the transitions along a stochastic trajectory. These observables capture some dynamical aspects of the trajectories like, current, entropy production, traffic, etc., and hence become important for out-of-equilibrium systems. The average of such jump observables can again be computed using just the stationary distribution of the state. However, all higher moments require the knowledge of the full trajectory, due to the dependence of the different jumps on the state. 
In the following, we will study both kind of observables in the CMOS inverter.

The CMOS inverter is operated at a voltage bias $\Delta V > 0$, and it relaxes to a non-equilibrium steady state (NESS). Being a one-dimensional system, the stationary state for the CMOS inverter can be exactly solved by taking $d_tP_{\rm{st}}(q,t)=0$ from Eqn.~\eqref{eq: invertermastereqn}. As shown in \cite{freitas2021stochastic}, this simplifies to the following recurrence relation,
\begin{equation}
    P_{\rm{st}}(q)=\frac{\alpha_p+\alpha_n\gamma e^{-(q-q_e)/q_T}}{\alpha_n+\alpha_p \gamma e^{q/q_T}}P_{st}(q-q_e),
    \label{eqn: exact_sol}
\end{equation}
where, 
\begin{eqnarray}
&\alpha_n =e^{(V_{\text{in}}-V_{\text{ss}})/nV_{\rm{T}}}  	\hspace{3em} \alpha_p=e^{(V_{\text{dd}}-V_{\text{in}})/nV_{\rm{T}}}\nonumber\\
&\gamma =e^{-(\Delta V+v_e)/(2V_{\rm{T}})}.\nonumber
\end{eqnarray}
This, combined with the normalization condition $\sum_q P_{\rm{st}}(q) = 1$, gives us the exact form for the stationary probability distribution. Hence, the stationary statistics of any state observable (a function of output charge) and the average of any jump observable can be exactly computed.  For example, the average steady state current through the pMOS transistor is
\begin{eqnarray}
 \langle I_{\rm{p}}\rangle = \sum_q P_{\rm{st}}(q)\left[ \lambda_+^\text{p}(q)- \lambda_-^\text{p}(q)\right].
 \label{eqn: av. current}
\end{eqnarray}
In the next section, we will first review the large deviation analysis of the statistics of state observables in the macroscopic limit, and solve it in the specific case of the CMOS inverter. We will compare it with the exact results above and also with the results obtained from the usual Gaussian approximation. We will show the limitations of the latter due to thermodynamic inconsistency. Then, we will consider the large deviation analysis for the current observables in both the macroscopic and long time limit. 

\section{Macroscopic Limit and Large Deviation Principle for state observables}
\label{sec: quasi}

For most electronic systems, analytically solving the master equation even for the stationary state is not feasible. One thus has to rely on numerical procedures, like spectral methods (diagonalization of the generator of the master equation), or stochastic simulation methods using the Gillespie algorithm \cite{gillespie1977exact}. While the former methods are limited to low-dimensional state spaces, the latter ones are computationally intensive.
However, analytical progress is possible by exploiting a low-noise macroscopic limit. In electronic circuits, this limit arises naturally under some scaling of the physical dimensions of the device.
Essentially, one enters the macroscopic regime when the elementary voltage $v_e=q_e/(2C_o)$ becomes negligible compared to all the other voltage scales in the system ($V_{\rm{dd}}$ and $V_{\rm{T}}$ in this case).
For example, when one increases the width $W$ of the conduction channel in a MOS transistor, both the capacitance $C_o$  and the specific current $I_0$ increase proportionally to $W$. Therefore, the elementary voltage $v_e^{-1} = q_e/(2C_0)$ decreases as $W^{-1}$ while the transition rates in the master equation increase as $W$.
Since the typical number of charges in the output conductor scales as $C_o$, it is natural to work with the voltage variable $v\sim q/(2C_o)$. Taking $v_e^{-1}$ itself as the scale parameter, the macroscopic limit corresponds to taking $v_e\to0$. We can hence define rescaled Poisson rates (still dependent of $v_e$) as $\omega_\rho(v,v_e)= v_e\lambda_\rho(v,v_e)$, such that $\omega_\rho(v) \equiv \lim_{v_e\to0}\omega_\rho(v,v_e)$ is well-defined. 

To study the macroscopic limit, it is convenient to rewrite the master equation in terms of the scaled rates as
\begin{equation}
\begin{split}
     d_tP(v,t)=\sum_\rho v_e^{-1}\bigg[&\omega_\rho(v-\Delta_\rho v_e ,v_e)P(v-\Delta_\rho v_e)\nonumber\\
     -\:&\omega_\rho(v,v_e)P(v,t)\bigg],
     \label{eqn: scaled_ME}
\end{split}
\end{equation}
and consider its series expansion in powers of the small parameter $v_e$ (Kramers-Moyal/system size expansion) \cite{van1992stochastic}.  To the lowest order in $v_e$ we obtain the continuity equation 
\begin{eqnarray}
 d_t P(v,t) &=& -\partial_v \left[\sum_\rho \Delta_\rho \omega_\rho(v) P(v,t)\right]\\
 &=&-\partial_v[\mu(v)P(v,t)] +\mathcal{O}(v_e)
 \label{eqn: deterministic}
\end{eqnarray}
where we have defined the deterministic drift $\mu(v)=\sum_\rho \Delta_\rho \: \omega_\rho(v)$. A solution of the equation above is $P(v,t)=\delta(v-x(t))$, where $x(t)$ is given by the dynamics $\dot{x}(t)=\mu(x(t))$. This is just the deterministic dynamics for $v$ in Eq.~\eqref{eqn: detkitchoff}.\par

\begin{figure*}[ht!]
\centering
\includegraphics[width=\textwidth]{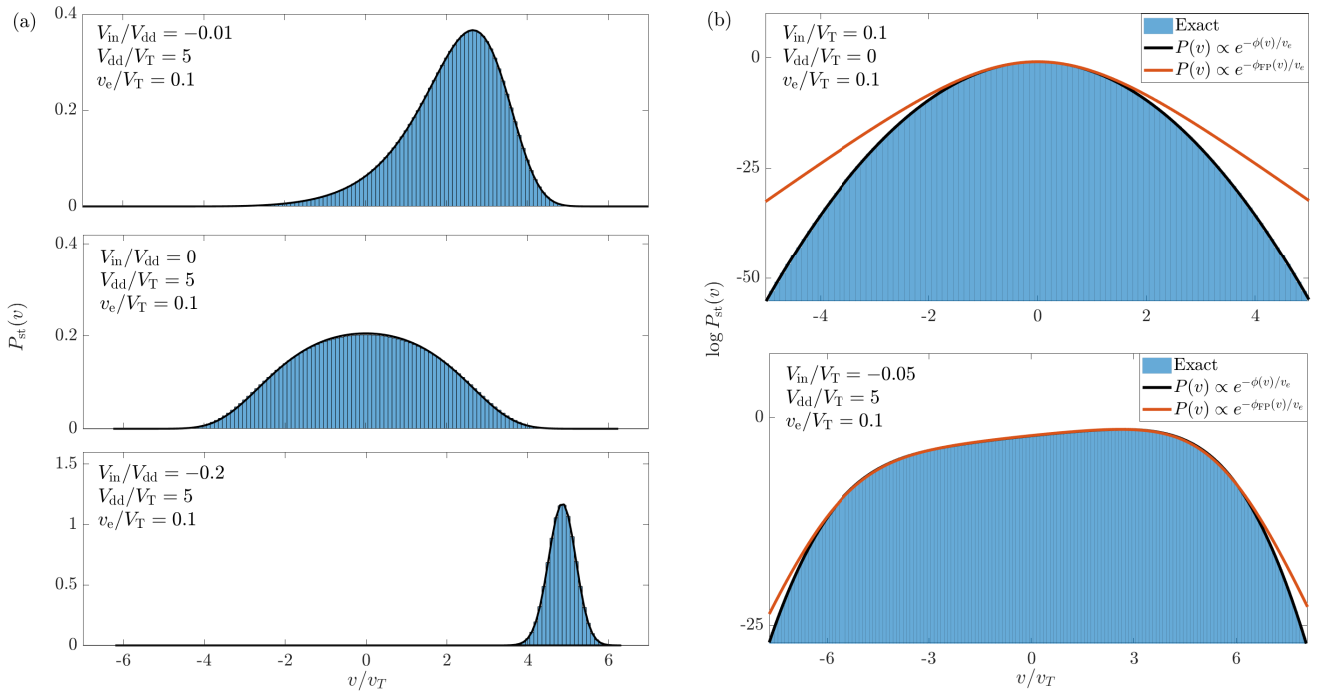}
\caption{(a): The exact steady state distribution of output voltage $v$ for the CMOS inverter obtained using the quasipotential given by Eq.~\eqref{eqn: quasipotential} (solid lines) is compared with the histogram from the exact solution (Eq.~\eqref{eqn: exact_sol}). (b): The stationary rate functions, the quasipotential $\phi$ and FP rate function $\phi_{\rm{FP}}$ along with the one computed from the exact solution $\log P_{\text{st}}(v)$ for an equilibrium and a non-equilibrium case.  }
\label{fig:Gaussian}
\end{figure*}

To the next order in $v_e$, we obtain the Fokker-Planck (FP) equation 
\begin{eqnarray}
     \partial_t P(v,t)=-\partial_v\left[\mu'(v)P(v,t)-\frac{1}{2}\partial_v(D(v)P(v,t))\right]
     \label{eqn:FP eqn}
\end{eqnarray}
involving a corrected drift term, 
\begin{eqnarray}
\mu'(v)&=&\sum_\rho \Delta_\rho [\omega_\rho(v) \hspace{-0.05cm}+\hspace{-0.05cm} v_e\partial_{v_e}\omega_\rho(v,0)]\hspace{-0.1cm\nonumber}\\
&=&\hspace{-0.1cm} \mu(v)+\sum_\rho v_e\Delta_\rho\partial_{v_e}\omega_\rho(v,0)
\end{eqnarray}
and the diffusion term, 
\begin{eqnarray}
D(v)=v_e\sum_\rho \omega_\rho(v).
\end{eqnarray} 
Since it is one-dimensional, this equation can be solved exactly, 
as shown in Appendix \ref{Appendix : FP solution}. Eq.~\eqref{eqn:FP eqn} is equivalent to the Langevin equations obtained by adding white noise sources to the deterministic dynamics (Eq.~\eqref{eqn: detkitchoff}), which is a common approach for the description of noisy non-linear systems \cite{gray2009analysis,nepal2005designing}. However, that approach is thermodynamically inconsistent. This can be seen by noting that for equilibrium settings (no powering bias, $V_\text{dd}=0$) the detailed balance condition is violated, i.e.,
\begin{eqnarray}
     2\mu'(v)\neq\partial_vD(v)-\beta D(v)\partial_v U(v)
     \label{eqn: FDT}
\end{eqnarray}
where $U(v)$ is the energy of the system. As a consequence, the steady state of the FP equation is not the Gibbs state $P_{\rm{eq}}(v)\sim e^{-U(v)/(k_BT)}$.
As we will see next, deviations from the correct equilibrium only occur for large fluctuations. However, these fluctuations are crucial to properly evaluate the probability of logical errors induced by thermal noise \cite{freitas2021reliability,hanggi1988bistability}. 

\par
Thus, the second-order truncation of the master equation expansion leads to systematic errors for rare events and also thermodynamic inconsistencies. 
An alternative method to study the macroscopic regime is provided by large deviations (LD) theory. 
The basis of this theory is the existence of an asymptotic form for the probability density of any intensive observable (voltage in this case), capturing the fact that fluctuations decay exponentially with the size of the system. This is known as the Large Deviation Principle. 
In our case, this principle is expressed as the existence of the limit $\phi(v,t) \equiv \lim_{v_e\to0}\left[-v_e\log P(v,t)\right]$, which leads to the LD ansatz 
\begin{eqnarray}
 P(v,t)\asymp e^{-v_e^{-1}[\phi(v,t)+o(v_e)]}.
 \label{eqn:ld ansartz}
\end{eqnarray}
Large deviation theory has been a powerful tool set in statistical physics, as the quantity $\phi(v,t)$ captures fluctuations beyond the central limit theorem. In the steady state, the rate function $\phi(v)$ is termed as quasipotential, by analogy with the Gibb's state in the equilibrium case.
Under the scaling conditions discussed above, we can see that the master equation accepts the LD ansatz by introducing Eq.~\eqref{eqn:ld ansartz} into  Eq.~\eqref{eqn: scaled_ME}, while keeping only the dominant terms in $v_e \to 0$.
We then obtain the following dynamical equation for the rate function
\begin{eqnarray}
 d_t \phi(v,t) =\sum_\rho \omega_\rho(v)\left[1-e^{\Delta_\rho\partial_v\phi(v,t)}\right].
\end{eqnarray}
Thus, the steady state quasipotential for the CMOS inverter satisfies
\begin{equation}
\begin{split}
 0 &= [\omega_-^\text{n}(v) + \omega_+^\text{p}(v)](e^{\partial_v \phi(v)}-1)\\ 
 &+[\omega_+^\text{n}(v) + \omega_-^\text{p}(v)](e^{-\partial_v \phi(v)}-1).
\end{split}
\end{equation}
Remarkably, the previous equation can be solved exactly, and one finds:
\begin{equation}
    \phi(x)=\frac{x^2}{2V_{\rm{T}}}+\frac{V_{\text{dd}}x}{V_{\rm{T}}}+V_{\rm{T}}\left[\text{Li}_2(x,V_{\rm{dd}})-\text{Li}_2(x,-V_{\rm{dd}})\right],
    \label{eqn: quasipotential}
\end{equation}
where $\text{Li}_2(x,V_{\rm{dd}}) = \text{Li}_2(-e^{(V_{\text{dd}}+x-(2/n) V_{\text{in}})/V_{\rm{T}}})$ is the poly-logarithm function of order 2.  One also notes that it provides a  thermodynamically consistent picture as in equilibrium ($V_{\rm{dd}}=0$), it gives back the correct thermodynamic potential $U(v)$.\par
To see the differences between the LD approach and the FP approach, we can also compute the large deviation rate function at the FP level. This motivates us to use the ansatz $P(v,t)\asymp \exp(-v_e^{-1}\phi_{\rm{FP}}(v,t))$, that when introduced in the FP equation leads to 
\begin{eqnarray}
 d_t \phi_{\rm{FP}}(v,t) = \mu(v)\partial_v \phi_{\rm{FP}}(v,t)+d(v)[\partial_v \phi_{\rm{FP}}(v,t)]^2
\end{eqnarray}
where we have defined $d(v)=D(v)/(2v_e)$. Note that the drift term now corresponds to just the deterministic part, $\mu(v)$, as the correction term only has a subexponential contribution. The corresponding steady state rate function is then obtained by solving $0=\mu(v)\partial_v \phi_{\rm{FP}}(v,t)+d(v)[\partial_v \phi_{\rm{FP}}(v,t)]^2$, which gives
\begin{eqnarray}
 \phi_{\rm{FP}}(v)&=&\int\left[-\frac{\mu(v')}{d(v')}\right]dv'.
\end{eqnarray}
The explicit form of the rate function $\phi_{\rm{FP}}(v)$ 
is given in Appendix.~\ref{Appendix : FP solution}. \par
All the results obtained so far are compared in Figure~\ref{fig:Gaussian}.
In Figure~\ref{fig:Gaussian}-(a), we compare the exact stationary distribution for the output voltage $v$  with the quasipotential solution (Eq.~\eqref{eqn: exact_sol}), for different $V_{\rm{in}}$ at fixed powering voltages $\Delta V$. One can see that the quasipotential is in good agreement with the exact solution, even for a finite system size ($v_e/V_{\rm{T}}=0.1$). 

In Figure~\ref{fig:Gaussian}(b), we compare the quasipotential with the stationary rate function from the FP equation,  $\phi_{\rm{FP}}$, for an equilibrium and a non-equilibrium case. We see that $\phi_{\rm{FP}}$ fails to capture the tails of the distribution, the rare fluctuations in the output voltage (Eq.~\eqref{eqn: FDT}). Thus, in addition to being the thermodynamically consistent approach, the LD quasipotential also captures the tails of the distribution for non-equilibrium cases.

\section{Spectral Methods for Current Statistics}
\label{sec: Spectral}
We will now study the statistics of the electric current through the CMOS inverter. The statistics of jump-like observables can provide insights into the fluctuations of the thermodynamic quantities like dissipation, entropy flow, etc.  The full counting statistics of currents have been previously explored in single electron systems, and have been used to verify the fluctuation theorems \cite{bagrets2003full,esposito2009nonequilibrium}. In this section, we will review the spectral methods to numerically compute the current statistics for a finite system size, and then use the limit of long times to characterize the typical fluctuations through large deviation functions \cite{lebowitz1999gallavotti,lazarescu2015physicist,touchette2011large}. In the following section, we will exploit both the long time limit and the large size limit.

Let us consider the stochastic trajectory $\{q(t)\}$ to be a particular realization of the dynamics given by Eq.~\eqref{eqn: ME_generator}, defined from some initial time $\tau=0$ up to a time $\tau=t$. It can be characterized by the initial state $q_0$ and the sequence of jumps $\{\rho_k\}$ along with their time stamps $\{\tau_k\}$, where the index $k$ is over all jumps. This information is encoded in the trajectory probability current, 
\begin{eqnarray}
 j_\rho(q,t)=\sum_{k}\delta[\rho,\rho_k]\delta[q,q_{t_k}]\delta(t-t_k)
\end{eqnarray}
where $q_t$ is the state immediately before the instant $t$. Then, the instantaneous electric current through a device $\rho$, is defined as $i_\rho(\tau) = q_e \sum_q [j_\rho(q,\tau)-j_{-\rho}(q+q_e\Delta_\rho,\tau)]$. Other current observables relevant in electronics like power, entropy production rate, etc., can be defined similarly.\par
To compute the statistics of such current observables, it is useful to define the time-integrated current observable $N_\rho(t)= \int_0^t d\tau\: i_\rho(\tau)=q_e\sum_{k}s(\rho_k)$ where $s(\rho_k)$ is the sign function when $\rho_k=\pm \rho$, else $0$. The statistics of $N_\rho(t)$ can be characterized by the moment generating function (MGF), defined as  $Z_\rho(\xi,t)=\langle e^{ \xi N_\rho(t)}\rangle$. It contains the same amount of information as the corresponding probability distribution $P(N_\rho,t)$, but $Z(\xi,t)$ is usually easier to compute. One can then recover $P(N_\rho,t)$ by inverting 
the MGF
\begin{eqnarray}
 P(N_\rho,t)=\int_{-\pi}^{\pi}\frac{d\xi}{2\pi}Z_\rho(i\xi,t)e^{-i\xi N_\rho}.
\end{eqnarray}
In the matrix representation, the evolution of $Z(\xi,t)$ is given by the `tilted' dynamics
\begin{eqnarray}
  Z_\rho(\xi,t) = \bra{1}e^{\hat{\mathcal{L}}_\xi t}\ket{P(0)}
  \label{eqn: spec_current_dynamics}
\end{eqnarray}
where $\bra{1}\equiv(1,1,...)$, $\ket{P(0)}$ is the initial probability vector and $\hat{\mathcal{L}}_\xi =\hat{\Gamma}_\xi(t)-\hat{\gamma}(t)$ is the tilted generator
with non-diagonal elements given by:
\begin{eqnarray}
       \hat{\Gamma}_\xi(t) &=&\sum_q\sum_{\rho'}\ket{q}\bra{q-q_e\Delta_{\rho'} }\lambda_\rho(q-q_e\Delta_{\rho'} ) e^{\xi s(\rho')}\nonumber.
\end{eqnarray}
Note that $\hat{\mathcal{L}}_\xi$ has an additional weight for the cross (jump) terms $\hat{\Gamma}_\xi(t)$ by $e^{\xi s(\rho')}$, where $s(\rho')$ is the sign function as defined before. One may also note that $\hat{\mathcal{L}}_\xi$ cannot be a generator for any Markov process as it does not preserve normalization, i.e., $\bra{1}\hat{\mathcal{L}}_\xi\neq0$.

The tilted dynamics of Eq.~\eqref{eqn: spec_current_dynamics} can be used to compute the full counting statistics (FCS) of the integrated current at any finite time. However, in practice, the exponential matrix evolution (Eq.~\eqref{eqn: spec_current_dynamics}) is not always feasible in the macroscopic limit (large state space) and one also has to be careful with the truncation of the state space for accurate results. In such cases, one could still get the dominant behaviour of the statistics at long times by looking at the largest eigenvalue of the tilted generator $\hat{\mathcal{L}}_\xi$. The MGF can be spectrally decomposed as, $Z_\rho(\xi,t)=\sum_j c_j(\xi) e^{\lambda_j(\xi)t}$, where $\lambda_j(\xi)$ are the eigenvalues of $\hat{\mathcal{L}}_\xi$  and the $c_j(\xi) $ are the coefficients of the expansion. The existence of unique maximal eigenvalue $\max[\lambda_j]$ of $\hat{\mathcal{L}}_\xi$ (Perron-Frobenius Theorem)\cite{chetrite2015nonequilibrium}, implies that the MGF satisfies
\begin{eqnarray}
\lim_{t\to\infty} \frac{\log Z_\rho(\xi,t)}{t} = \lim_{t\to\infty}\frac{1}{t}\log\langle e^{\xi N_\rho(t)}\rangle = S_\rho(\xi),
\label{eqn: spec_LD}
\end{eqnarray}
where $S_\rho(\xi)= \max[\lambda_j]$ is the scaled cumulant generating function (SCGF) for the integrated current through the device $\rho$. This is equivalent to having a large deviation form for the MGF at long times. The SCGF enjoys a fluctuation symmetry $S_\rho(\xi)=S_\rho(-\xi-\beta q_e \Delta V)$, where $\Delta V$ is the potential difference across the device (App.~\ref{Appendix : Fluctuation symmetry}). This symmetry provides a thermodynamic consistency check for the various approximations discussed below. The G\"artner-Ellis theorem implies that the time-intensive empirical current $O\coloneqq I_\rho(t)=N_\rho(t)/t=i$ also satisfies a large deviation principle, where the large scale parameter is $t$. So asymptotically, the probability density has an exponential form, $P(I_\rho(t)=i)\sim e^{-t\phi(i)}$, where $\phi(i)$ is the rate function for $i$. Also, $S(\xi)$ is related to $\phi(i)$ by the Legendre-Fenchel transform 
\begin{eqnarray}
 S_\rho(\xi)= \max_{i}\left[ \xi i - \phi_\rho(i)\right].
 \label{eqn: legendre-fenchel}
\end{eqnarray}
Since the current $I_\rho$ depends on the integration time $t$ of the measurement, the normalized cumulants from the SCGF $S_\rho(\xi)$ can be obtained as follows,
 \begin{eqnarray}
 \langle\langle I_\rho^n\rangle\rangle t^{n-1}= \; \partial_\xi^n S_\rho(\xi)|_{\xi=0},
 \label{eqn: cumulants_current}
\end{eqnarray}
where $ \langle\langle I_\rho^n\rangle\rangle$ is the $n^{th}$ cumulant of the empirical current $I_\rho$. 
The large deviations principle discussed above is a consequence of taking the large time limit, and it does not involve taking the macroscopic limit of the devices. Compared to long-time Gillespie simulations, the spectral method has the advantage of computing just the largest eigenvalue of the biased generator. However, for electronic systems it remains as a numerical tool, which also becomes expensive in the macroscopic limit. In the next section, we will show how analytical progress can be made by first taking the macroscopic limit and then taking a large time limit for the current statistics.

\section{Macroscopic limit and Large Deviation Principle for Current Statistics}
\label{sec: current_macro}

The spectral methods for current statistics do not exploit the macroscopic limit that was introduced in section III for state observables. As a consequence, they are limited to situations in which the state space can be truncated to a manageable number of dimensions, so that the generator of the master equation can be diagonalized, or at least its dominant eigenvalue can be computed. However, the macroscopic limit can also be considered for the current statistics using the path integral representation of the stochastic process. Path integral tools, first introduced to stochastic systems by Wiener \cite{wiener1923differential}, have been applied to study a wide range of problems in non-equilibrium systems \cite{herman1982path,lehmann2000surmounting,roldan2017path,van1992stochastic,gardiner2009stochastic,wio2013path}. They allow to express the probability of a given stochastic trajectory in terms of an action. For a system displaying the macroscopic limit, the scaling of the action with respect to the scale parameter implies the emergence of typical trajectories. For example, in the case of the inverter, the probability of a trajectory $\{v(t)\}$ can be expressed using the Martin-Siggia-Rose (MSR) representation \cite{martin1973statistical,peliti1985path,tomassothesis} as
\begin{equation}
    P[\{v(t)\}] = \int \mathcal{D}p \: e^{v_e^{-1} \int_0^t d\tau \left[ -p(\tau) \dot v(\tau) + 
    H[v(\tau), p(\tau)]\right]},
    \label{eq:basic_path_int}
\end{equation}
where the argument of the exponential, the action, is expressed to dominant order in the scale parameter $v_e^{-1}$ in terms of a Hamiltonian
\begin{equation}
    H[v,p]=\sum_{\rho}\left(e^{\Delta_{\rho} \: p }-1\right)
    \omega_{\rho}(v).
\end{equation}
In Eq.~\eqref{eq:basic_path_int}, $\int \mathcal{D}p$ is a path integral over all the possible configurations of the auxiliary field $p(t)$, playing the role of a conjugated momentum. Extremizing the action, one obtains the typical trajectories, following Hamilton's equations of motion in the extended space $(v,p)$:
\begin{eqnarray}
          \dot{v}(\tau) &=& \sum_\rho\Delta_\rho e^{\Delta_\rho p(\tau) }\omega_\rho(v(\tau)),\label{eqn: eom_v}\\
    \dot{p}(\tau) &=&-\sum_\rho\left( e^{\Delta_\rho p(\tau)}-1\right)\partial_v\omega_\rho(v(\tau)).
    \label{eqn: eom_p}
\end{eqnarray}
Note that setting $p=0$, one recovers the deterministic equations of motion in Eqs.~\eqref{eqn: eom_v} and \eqref{eqn: eom_p}. 
The stationary state (quasipotential) can be computed from the trajectories which nullify the Hamiltonian (i.e., $H[v,p]=0$) in Eq.~\eqref{eq:basic_path_int} in the long time limit. The details of the exact derivation of the quasipotential and the FP approximation from the path integral approach can be found in the Appendix.~\ref{app: LD_quasi}.\par
The generating function for current statistics introduced in the previous section also accepts a path integral representation. Indeed, again to the dominant order in $v_e^{-1}$, we have:
\begin{equation}
    Z_\rho(\xi, t) \!=\! \! \int\!\! \mathcal{D}v \mathcal{D}p \:\: e^{v_e^{-1} \! \!\int_0^t d\tau \left[ -p(\tau) \dot v(\tau) + 
    H_\xi[v(\tau), p(\tau)]\right]} P_0(v(0)),
    \label{eq:current_path_int}
\end{equation}
where $P_0(v(0)) \propto e^{-v_e^{-1}\phi_0(v_0) + o(v_e^{-1})}$ is an initial probability distribution, and $H_\xi(v,p)$ is the \emph{biased} Hamiltonian
\begin{eqnarray}
    H_\xi[v,p]=\sum_{\rho}\left(e^{\Delta_{\rho} p +s(\rho)\xi}-1\right)\omega_{\rho}(v).
    \label{eqn: biased_H}
\end{eqnarray}
The above Hamiltonian is biased by the counting field $\xi$ corresponding to the current observable, similar to that of the tilted generator (Eq.~\eqref{eqn: spec_current_dynamics}) introduced in the previous section. The macroscopic limit $v_e \to 0$ can be now exploited to perform a saddle point approximation of the path integral in Eq.~\eqref{eq:current_path_int}, where only the most probable trajectory $(v_\xi(t), p_\xi(t))$ for each value of the bias $\xi$ is considered:
\begin{equation}
    Z_\rho(\xi, t) \simeq \max_{v_\xi(t), p_\xi(t)} e^{v_e^{-1} \int_0^t d\tau \left[ -p_\xi(\tau) \dot v_\xi(\tau) + 
    H_\xi[v_\xi(\tau), p_\xi(\tau)]\right]}.
    \label{eqn: Z_finitet}
\end{equation}
In the previous equation, the optimization must be performed among all the trajectories $(v_\xi(t), p_\xi(t))$ extremizing the action, or equivalently satisfying the corresponding Hamiltonian dynamics:
\begin{eqnarray}
    \dot{v_\xi}&=\partial_p H_\xi(v_\xi,p_\xi) \qquad
    \dot{p_\xi}&=-\partial_v H_\xi(v_\xi,p_\xi)
    \label{eqn:Z_eom}
\end{eqnarray}
with boundary conditions
\begin{eqnarray}
p(0)=\frac{\partial}{\partial v_0}\phi_0(v_0), \qquad p(t)=0.     
\label{eqn: boundary_cond_current}
\end{eqnarray}
Solving the above non-linear coupled equations of motion analytically can be difficult. However, the optimization over all the trajectories can be avoided if we are only interested in the current statistics for long integration times $t$. In that case, the optimal trajectories are the ones which pass extremely close to the fixed points $(v_\xi^*, p_\xi^*)$ of the biased dynamics generated by $H_\xi$, in addition to satisfying the boundary conditions. Then, the kinetic term $p\dot{v}$ vanishes in the bulk of long trajectories, and the only time extensive contribution to the integral in Eq.~\eqref{eqn: Z_finitet} is simply $t H_\xi(v_\xi^*, p_\xi^*)$. In this way, we find that the scaled cumulant generating function (SCGF) for the current through a device $\rho$ is given by,
\begin{eqnarray}
     S_\rho(\xi)=\lim_{t\to\infty}\frac{1}{t}\ln Z_\rho(\xi,t)=v_e^{-1}\max_{v^*_\xi,p^*_\xi}\left\{ H_\xi[v^*_\xi,p^*_\xi]\right\}
     \label{eqn: SCGF_Hamiltonian}
\end{eqnarray}
This is the central result of the path integral formalism that will be used in the following. 
As shown in Appendix.~\ref{Appendix : Fluctuation symmetry}, the Lebowitz-Spohn fluctuation symmetry $S_\rho(\xi)=S_\rho(-\xi-\beta q_e \Delta V)$ is satisfied by Eq. \eqref{eqn: SCGF_Hamiltonian}. Note that the role of the eigenvalue with the largest real part in the spectral method is played by the scaled biased Hamiltonian, $H_\xi[v^*_\xi,p^*_\xi]/v_e$.

In the path integral method discussed above, one first takes the macroscopic limit, $v_e\to 0$, and only then the $t\to\infty$ limit is taken. Whereas in spectral methods, the long time limit is taken first for any finite system size (which could also be taken to be large). These two limits not always commute and may give different insights into the current statistics, for example in systems with long-lived metastable states. In such systems, taking the macroscopic limit first leads to ergodicity breaking. Then, the dynamics gets stuck near one of the metastable states and the distribution for the current is peaked around the deterministic value corresponding to that state (Appendix.~\ref{Appendix: Minimum rate function}). In contrast, ergodicity is preserved by taking the long time limit first for a finite size, and the dynamics explores the full state space. Then the
most probable value for the integrated current is the average between the values corresponding to each metastable state (weighted by their respective lifetimes) \cite{jack2020ergodicity}. 
\par

\subsection{Application to the CMOS inverter}

We will now apply the central result Eq.~\eqref{eqn: SCGF_Hamiltonian} to the CMOS inverter. If one is interested in the statistics of the electric current through the pMOS transistor $I_p$, the corresponding biased Hamiltonian reads:
\begin{flalign}
H_\xi^{\rm{pMOS}}[v,p] &= (e^{p}\!-\!1)\: \omega_-^\text{n}(v)+(e^{-p}\!-\!1)\:\omega_+^\text{n}(v)\\
&+(e^{p+\xi}\!-\!1)\:\omega_+^\text{p}(v)+(e^{-p-\xi}\!-\!1)\:\omega_-^\text{p}(v).\nonumber
\end{flalign}
Similarly, the biased Hamiltonian for the electric current through the nMOS transistor $I_n$, is given as
\begin{flalign}
H_\xi^{\rm{nMOS}}[v,p] &=(e^{p}\!-\!1)\:\omega_+^\text{p}(v)+(e^{-p}\!-\!1)\:\omega_-^\text{p}(v)\\
&+ (e^{p-\xi}\!-\!1)\:\omega_-^\text{n}(v)+(e^{-p+\xi}\!-\!1)\:\omega_+^\text{n}(v).\nonumber
\end{flalign}
These two Hamiltonians are related by the following symmetry:
\begin{equation}
H^{\rm{pMOS}}_\xi[v,p]= H^{\rm{nMOS}}_\xi[v,p+\xi]   
\label{symmetry_pn}
\end{equation}
According to this relation, there is a one-to-one mapping between the fixed points of the two Hamiltonians, and their values at the corresponding fixed points match. As a consequence, the currents through both transistors share the same SCGF, and therefore the same statistics in the long time limit. This is naturally understood as follows. Conservation of charge implies that the net number of charges $N_\text{nMOS}$ and $N_\text{pMOS}$ transported by the two transistors during time $t$ can only differ by the charge fluctuation $\Delta q = q(t)-q(0)$. At steady state, $\Delta q/t \to 0$ for long $t$, since $\Delta q$ is bounded, and therefore 
$N_\text{nMOS}/t \simeq N_\text{pMOS}/t$.

\begin{figure}[h!]
\centering
\includegraphics[trim=100 100 500 0, clip,width=0.5\textwidth]{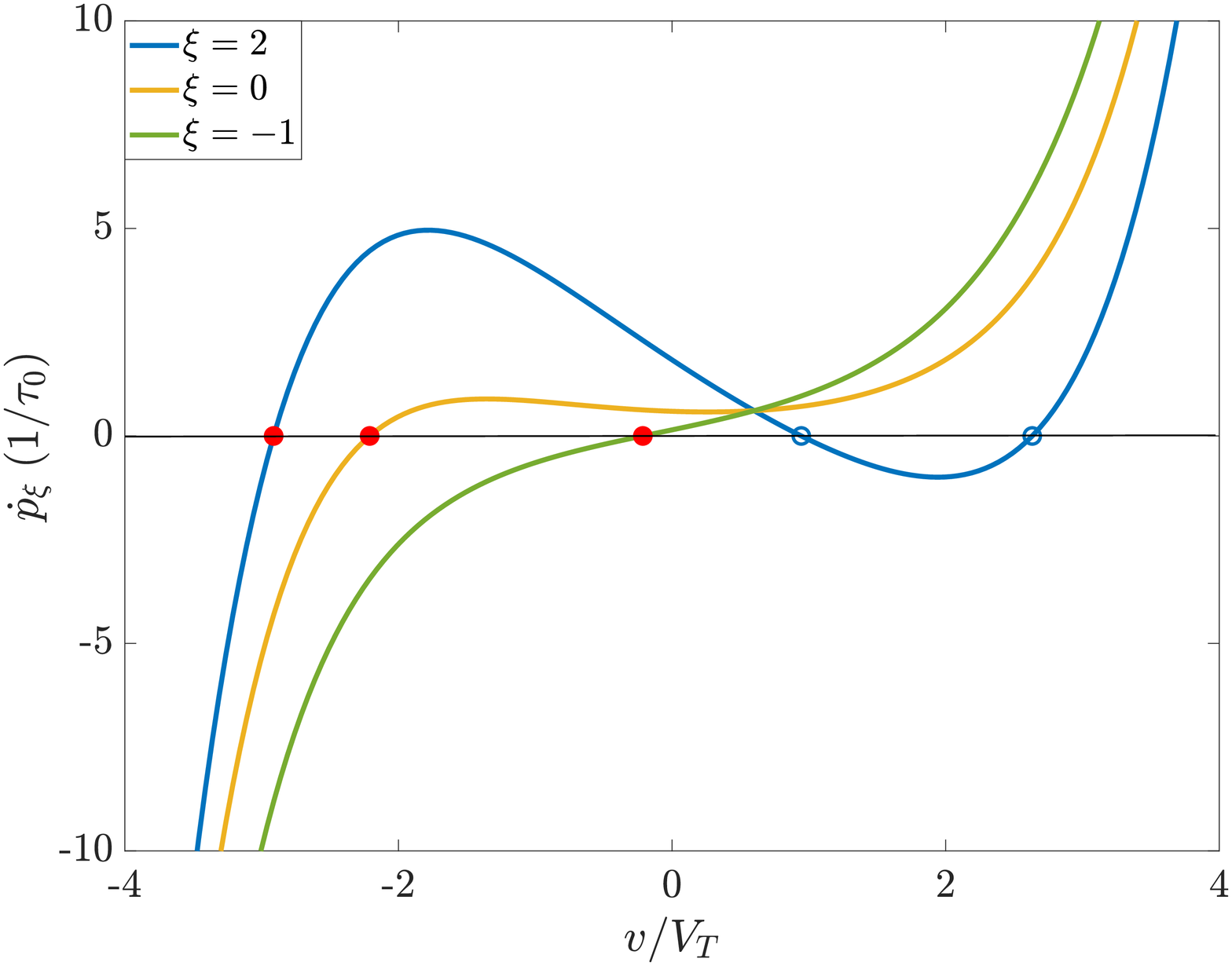}
\caption{ $\dot{p}_\xi(v)$ (in  units of $\tau_0^{-1}$, with $\tau_0=(q_e/I_0)\exp[V_{\rm{th}}/(nV_{\rm{T}})]$) for the nMOS transistor at different values of $\xi$, with the dominant fixed point marked in red. One can see that there are 3 different roots for $\xi=2$, with the leftmost and rightmost roots corresponding to hyperbolic fixed points. The middle fixed point corresponds to a center type, which cannot be reached given the boundary conditions. The leftmost root in each curve
is the one maximizing $H_\xi^{\rm{nMOS}}[v^*_\xi,p^*_\xi]$.
Parameters : $V_{\text{dd}}/V_{\rm{T}}=3, V_{\text{in}}/V_{\text{T}}=0.3$. }
\label{fig: pdot}
\end{figure}

To proceed, we need to compute the fixed points of $H_\xi^\text{(p/n)MOS}[v,p]$. Since $H_\xi[v,p]$ is a convex function of $p$ for any fixed value of $v$, we first compute the minimum $p_\xi^*(v)$ from the nullcline  corresponding to $\dot{v}_\xi=0$. The nullclines for $\dot{v}_\xi$ of both devices are given as 
\begin{eqnarray}
p^*_{\rm{pMOS}}(v) &=&\frac{1}{2}\ln\left[\frac{e^{-\xi}\:\omega_-^\text{p}(v)+\omega_+^\text{n}(v)}{e^\xi\:\omega_+^\text{p}(v) + \omega_-^\text{v}(v)}\right]\quad ;\nonumber\\
\quad p^*_{\rm{nMOS}}(v)&=&\frac{1}{2}\ln\left[\frac{\omega_-^\text{p}(v)+e^{\xi}\:\omega_+^\text{n}(v)}{\omega_+^\text{p}(v) + e^{-\xi}\:\omega_-^\text{n}(v)}\right].
\label{eqn:p(v)}
\end{eqnarray}
Then the fixed points $v^*_\xi,p^*_\xi$ can be obtained numerically after injecting the above solution $p^*_\xi(v)$ and finding the roots of the equation $\dot{p}_\xi(v)=-\partial_v H_\xi(v,p^*_\xi(v))=0$ for each value of $\xi$. In Figure \ref{fig: pdot}, we plot $\dot{p}_\xi(v)$ for the nMOS current at different values of $\xi$. One can see that there can be multiple roots of the above equation ($\xi=2$), but we need to consider only those fixed points which satisfy the boundary conditions. Since the dynamics is Hamiltonian, the fixed points can have only marginal stability and the only possible  types in 2D are center and hyperbolic fixed points  \cite{strogatz2018nonlinear}. We can remove the ones of center type, as they do not satisfy the boundary conditions. Assuming that the rest of the hyperbolic fixed points are compatible with the boundary conditions, we choose the one that maximizes $H_\xi[v^*_\xi,p^*_\xi]$. \par

\subsection{Diffusive approximation for current statistics}

In the path integral formalism, the diffusive approximation (in which the Poisson noise is approximated by Gaussian noise) corresponds to expanding the Hamiltonian $H[v,p]$ to second order in $p$ (see Appendix.~\ref{app: LD_quasi}). This is analogous to the second-order expansion of the master equation (Eq. \eqref{eqn: scaled_ME}) in $v_e$, leading to the Fokker-Planck equation in Eq. \eqref{eqn:FP eqn}. We now apply that approximation to the biased Hamiltonian giving the current statistics for the underlying Langevin dynamics. Thus, expanding Eq. \eqref{eqn: biased_H} to second order in $p$, we obtain:
\begin{flalign}
H_\xi^{G}[v,p] &=\sum_\rho\left(e^{s(\rho)\xi}\!-\!1\right)\omega_\rho(v)+p\sum_\rho\Delta_\rho\: \omega_\rho(v)\:e^{s(\rho)\xi}\nonumber\\
&+\frac{p^2}{2}\sum_\rho\omega_\rho(v)\:e^{s(\rho)\xi}+\mathcal{O}(p^2).
\label{eqn: biasedH_gauss1}
\end{flalign}

However, the above diffusive approximation breaks the symmetry in Eq.~\eqref{symmetry_pn} between the two transistors, and hence the approximate long time current statistics for the pMOS and nMOS transistors are different, which is unphysical. Intuitively, this is due to the fact that the diffusive approximation fails to correctly describe large, non-Gaussian fluctuations. Since only Gaussian fluctuations are expected to be correctly described, we must only consider the first two terms of the expansion in $\xi$ of the SCGF.
Therefore, the appropriate way to perform the diffusive approximation 
for the current statistics is to expand the biased Hamiltonian $H_\xi[v,p]$ to second-order both in $p$ and $\xi$. Doing that, we obtain:
\begin{flalign}
H_\xi^{G}[v,p] &=\sum_\rho[\Delta_\rho p+s(\rho)\xi]\:\omega_\rho(v)\nonumber\\
&+\frac{1}{2}\sum_\rho[\Delta_\rho p+s(\rho)\xi]^2\:\omega_\rho(v)+o(p^2,\xi^2),
\label{eqn: biasedH_gauss2}
\end{flalign}
Following the same procedure as before for obtaining the fixed points, we first get the equations for the nullclines in $p$ for both devices,
\begin{flalign}
p^*_{\rm{pMOS}}(v) &= \frac{\mu(v)-\xi\left(\omega_p^+(v)+\omega_p^-(v)\right)}{\sum_\rho \omega_\rho}\nonumber\\
p^*_{\rm{nMOS}}(v)&=\frac{\mu(v)+\xi\left(\omega_n^+(v)+\omega_n^-(v)\right)}{\sum_\rho \omega_\rho}\nonumber.\nonumber
\end{flalign}
Using the above solutions, we numerically compute the fixed points of this Hamiltonian to obtain the SCGF. Similar to the state observables, the Gaussian approximation remains thermodynamically inconsistent, as it fails to capture the fluctuation symmetry $S_\rho(\xi)=S_\rho(-\xi-\beta q_e \Delta V)$ (Appendix.~\ref{app: LD_current}).
\par
\begin{figure}[h!]
\centering
\includegraphics[trim=70 25 450 0, clip,width=0.5\textwidth]{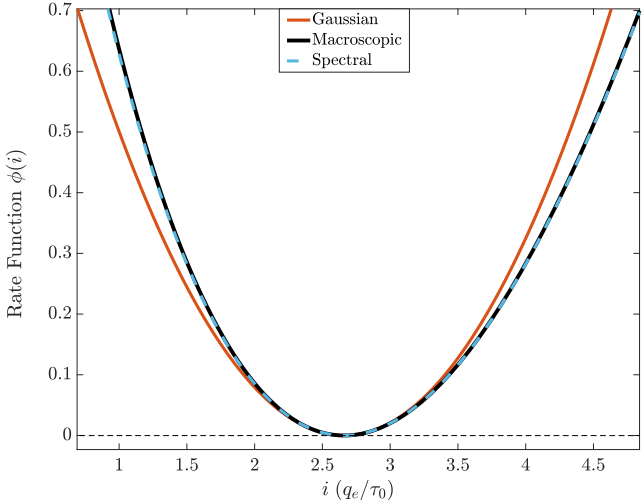}
\caption{The rate function $\phi(i)$ obtained by taking the Legendre-Fenchel transform of $S_\rho(\xi)$ given by spectral method (dashed light blue), macroscopic limit (solid black) and Gaussian approximation (solid orange) of the biased Hamiltonian.
    Parameters : $V_{\text{dd}}/V_{\rm{T}}=2, V_{\text{in}}/V_{\text{T}}=1, v_e/V_{\rm{T}}=0.1$. }
\label{fig: Ratefunc_current}
\end{figure}

The SCGF $S_\rho(\xi)$, obtained from the full Hamiltonian $H_\xi[v,p]$  in Eq. \eqref{eqn: biased_H} or the Gaussian Hamiltonian 
$H^G_\xi[v,p]$ in Eq. \eqref{eqn: biasedH_gauss2}, can be
transformed to obtain the corresponding rate functions ${\phi(I_{\rm{p/nMOS}}=i)}$ for the empirical currents (via the inverse of the Legendre-Fenchel transform in Eq. \eqref{eqn: legendre-fenchel}). These two rate functions are compared 
in Figure~\ref{fig: Ratefunc_current}, where we also plot the exact rate function obtained using the spectral method. 
Similar to the case of state observables, the large deviation approach captures the current statistics even for a finite size $v_e/V_{\rm{T}}=0.1$.
As expected, we see that the Gaussian approximation correctly captures the curvature around the most probable value, but fails to describe the tails since the full distribution is not Gaussian.
\par

\begin{figure*}[ht!]
\centering
\includegraphics[trim=0 120 0 90, clip,width=\textwidth]{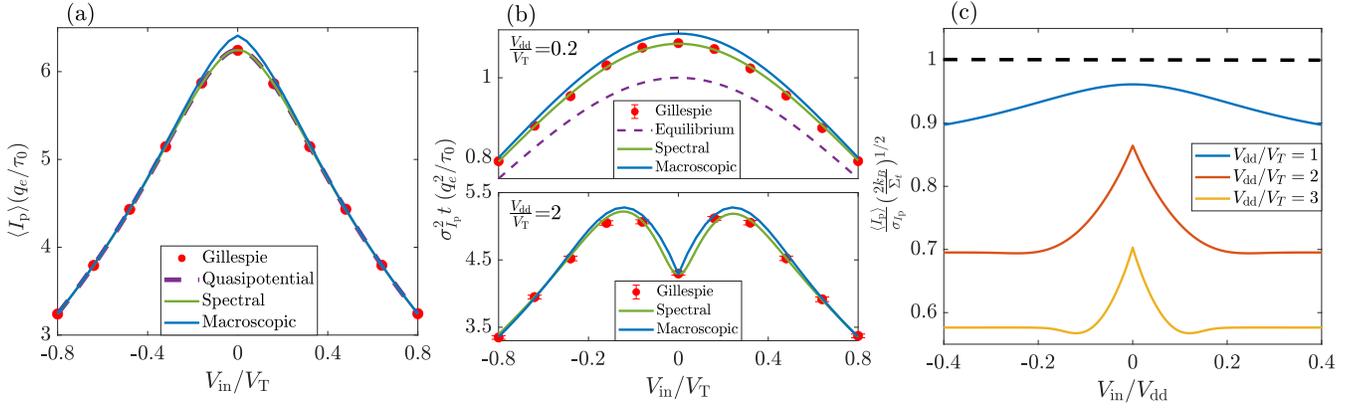}
\caption{(a) Average Current $\langle I_p \rangle$ through the pMOS transistor as a function of input voltage $V_{\rm{in}}/V_{\rm{T}}$ at a fixed powering voltage $V_{\rm{dd}}$ (symmetric case). The curves are obtained using the macroscopic limit, quasipotential estimate, and the spectral method, and compared with the Gillespie simulation results. Parameters : $V_{\rm{dd}}/V_{\rm{T}}=2,\, v_e/V_{\rm{T}}=0.1,\, t/\tau_0=10^3$. (b) The scaled variance ($t\sigma^2_{I_p}$) of the current through the pMOS transistor as a function of $V_{\rm{in}}/V_{\rm{T}}$ for the two cases : near equilibrium, $V_{\rm{dd}}/V_{\rm{T}}=0.2$ (upper), and far from equilibrium, $V_{\rm{dd}}/V_{\rm{T}}=2$ (lower). 
For $V_{\rm{dd}}/V_{\rm{T}}=0.2$, we also plot the equilibrium macroscopic result $\sigma_I^2 t|_{v_e\to0,V_{\rm{dd}}\to0}= 2\Omega$ (Eq.~\eqref{eqn: eqbrm_conductivity}), to show the similar qualitative behaviour.  Parameters : $ v_e/V_{\rm{T}}=0.1,t/\tau_0=5*10^3$. (c) The precision of current scaled by the entropy production (computed using the macroscopic limit) is plotted for different powering voltages $V_{\rm{dd}}/V_{\rm{T}}$, and compared with the TUR bound Eqn.~\eqref{eqn: TUR} (dashed line).}
\label{fig:av_var_current}
\end{figure*}

\section{Current Statistics for the CMOS Inverter}
\label{sec: CMOS_current}

Having established the different methods, and discussed their accuracy and limitations, we will now study the current statistics of the CMOS inverter for different parameters. We begin by analyzing the mean value of the current for different values of the input voltage $V_\text{in}$. In the first place, we note that the mean value obtained from the SCGF of Eq.~\eqref{eqn: SCGF_Hamiltonian} is just the deterministic value. For example, the average current through the pMOS transistor computed from $S_p(\xi)$ is
\begin{eqnarray}
\langle I_{\rm{p}}\rangle &= &\frac{\partial S_p(\xi)}{\partial\xi}\bigg|_{\xi=0} = \frac{\partial [H_\xi^{\rm{pMOS}}[v^*_\xi,p^*_\xi]/v_e]}{\partial\xi}\bigg|_{\xi=0}\nonumber\\
&=&\lambda_+^\text{p}(v_{\text{det}},0)-\lambda_-^\text{p}(v_{\text{det}},0)=I_{\rm{p}}(v_{\text{det}}).
\label{eqn: current_inverter_mean}
\end{eqnarray}
This is correct since the SCGF was obtained in the limit $v_e \to 0$. However, for any finite 
value of $v_e$, one can get a more accurate estimate of the average current by using the steady state 
quasipotential in Eq. \eqref{eqn: quasipotential}:
\begin{eqnarray}
 \langle I_{\rm{p}}\rangle = \int_{-\infty}^{\infty}dv \frac{1}{N}e^{-v_e^{-1}\phi(v)}\left[ \lambda_+^\text{p}(v,v_e)- \lambda_-^\text{p}(v,v_e)\right],
 \label{eqn: av_i_quasi}
\end{eqnarray}
where $N= \int  \:dv \: \exp(-\phi(v)/v_e)$ 
is the normalization constant.
Since the quasipotential $\phi(v)$ matches the state distribution even 
for finite $v_e$ (Figure~\ref{fig:Gaussian}), the average current computed in this way correctly incorporates contributions of the fluctuations that are not captured by Eq. \eqref{eqn: current_inverter_mean}. 
This also shows that the convergence of the large deviation approximation done at the state level
differs from that done at the current level. However, as already mentioned, 
higher order cumulants of the current at
long times cannot be computed just from the steady state distribution, and one needs to employ
the SCGF. In Figure \ref{fig:av_var_current}-(a), we show the average value $\mean{I}$ through the inverter as a function of $V_\text{in}$ for a fixed $V_\text{dd}$, computed using the different methods discussed until now.
We see that Gillespie simulations and the spectral 
methods give the same results, as they should, and that they also match the average based on the quasipotential (Eq. \eqref{eqn: av_i_quasi}). We also see that there are small deviations between those results and 
the large deviations estimation (Eq. \eqref{eqn: current_inverter_mean}) for small values of $V_\text{in}$, due to finite $v_e$ effects. 
The qualitative shapes of the curves are easily understood by recalling the transfer function of the inverter in Figure \ref{fig: CMOS inverter}-(b). We see that as we increase $|V_\text{in}|$, the output voltage approaches one of the two sources, which minimizes the voltage bias across the more conductive transistor, and maximizes the voltage bias across the less conductive transistor. 
The current is thus dominated by the less conductive transistor. Since the conductivity of the (n/p)MOS transistor scales as $e^{\pm V_\text{in}/V_{\rm{T}}}$, we see that the average current through the inverter scales as $e^{-|V_\text{in}/V_{\rm{T}}|}$ and is maximized at $V_\text{in} =0$.

We now turn to the analysis of the variance $\sigma^2_I = \mean{(I-\mean{I})^2}$. For the case of pMOS transistor, it is obtained from the SCGF at the macroscopic limit as:
\begin{eqnarray}
\sigma^2_{I_{\rm{p}}} \!=\! \frac{1}{t}\frac{\partial^2 S_p(\xi)}{\partial\xi^2}\bigg|_{\xi=0}\!=\!\frac{1}{t} \frac{\partial^2 [H_\xi^{\rm{pMOS}}[v^*_\xi,p^*_\xi]/v_e]}{\partial\xi^2}\bigg|_{\xi=0},
\label{eqn: current_inverter_var}
\end{eqnarray}
In Figure \ref{fig:av_var_current}-(b) we compare the scaled variance $\sigma^2_I\,t$ computed in this way with the 
spectral method and Gillespie simulations. We see that the macroscopic result matches well with 
exact results, even for system sizes involving just tens of
electrons ($V_{\rm{dd}}/v_e=0.1$), as was the case for state observables. This confirms the 
practical value of the large deviations techniques to compute the current fluctuations in non-linear 
electronic circuits, since Gillespie and spectral methods quickly become impractical for large state 
spaces. Close to the equilibrium (i.e., for low $V_\text{dd}$), the variance is also maximized at $V_{\rm{in}}=0$, like the average current. However, we see that for larger values of $V_\text{dd}$, there are non-trivial 
features in the variance and that it is maximized at $|V_\text{in}| \neq 0$. 
The decaying behaviour at large input voltages can 
be again understood from the fact that the conduction is dominated by the less conductive transistor.
We can gain intuition about the behaviour close to equilibrium in the following way. Considering first the equilibrium case, one can show that the variance is given as $\sigma_I^2 t = 2\, \Omega$, where $\Omega=\frac{d \langle I_{p/n} \rangle}{d (2V_{\rm{dd}})}\Big|_{V_{\rm{dd}}\to0}$ is the conductivity of the inverter. This result is a consequence of the fluctuation symmetry in the current statistics \cite{saryal2019thermodynamic}.
In the deterministic limit, we can compute this conductivity for the case of $n=1$, 
obtaining
\begin{equation}
\Omega=\frac{I_0}{2q_e V_{\rm{T}}} e^{-\frac{V_{\rm{th}}}{V_{\rm{T}}}}
\left(\!\cosh\left(\!\frac{V_{\rm{in}}}{V_{\rm{T}}}\!\right)
-\sinh\left(\!\frac{V_{\rm{in}}}{V_{\rm{T}}}\!\right)\tanh\left(\!\frac{V_{\rm{in}}}{V_{\rm{T}}}\!\right)\!\right)
\label{eqn: eqbrm_conductivity}
\end{equation}
The scaled variance computed using the above equation is plotted with a dashed line in Figure~\ref{fig:av_var_current}-(b), and one finds a similar qualitative behaviour also for small $V_{\text{dd}}$.

\begin{figure*}[ht!]
\centering
\includegraphics[scale=0.6]{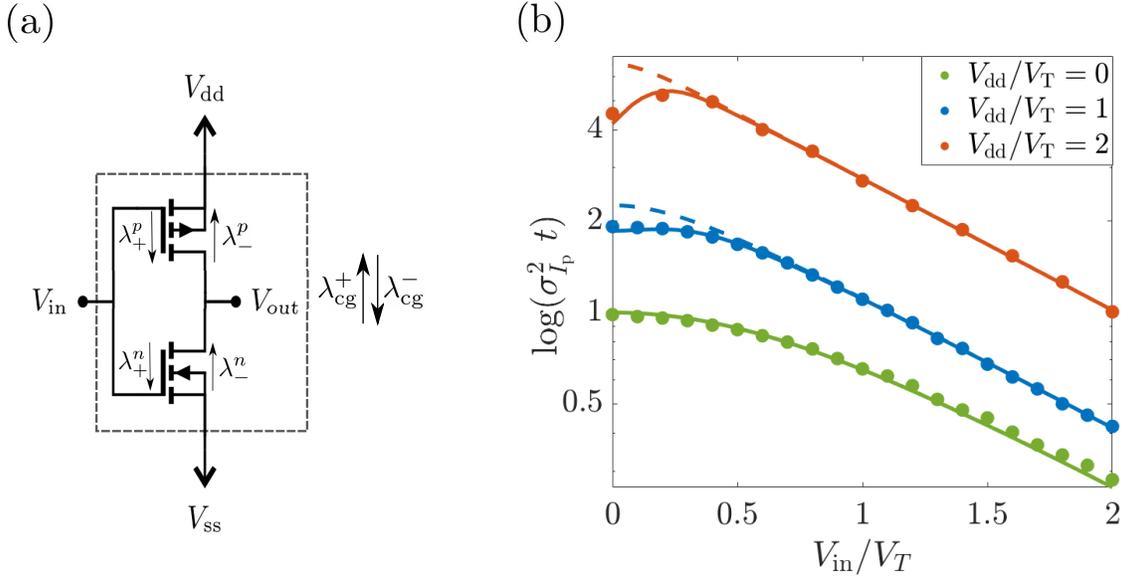}
\caption{The scaled variance of the current through the pMOS transistor $\sigma_{I_{\rm{p}}}^2\,t$, computed from the large deviation techniques (solid) and the coarse-grained dynamics (dashed), are plotted for different $V_{\rm{dd}}/V_{\rm{T}}$. The coarse-grained estimate for the variance is computed using the average current $\langle I \rangle$, using the exact solution of the stationary distribution (Eq.~\eqref{eqn: av. current}). At equilibrium ($V_{\rm{dd}}=0$), it is computed using the conductivity (Eq.~\eqref{eqn: eqbrm_conductivity}).
Parameters : $v_e/V_{\rm{T}}=0.1$. }
\label{fig:coarse-graining}
\end{figure*}

The non-monotonous behaviour of the variance for large values of $V_\text{dd}$ is harder to describe analytically. However, it is interesting to note that the nontrivial features (the maxima away from $V_\text{in}=0$) do not seem to be particular to the CMOS inverter, but a general feature of systems composed of two competing conduction channels affecting the same degree of freedom, with different timescales. We support this claim by showing that the same features are analytically reproduced using a simple and exactly solvable 2-state Markov jump process involving two channels, where we vary just the relative kinetic 
activity of the channels at fixed channel affinities (See Appendix \ref{Appendix: 2 state}). \par

\subsection{Thermodynamic uncertainty relations}
As shown in the previous section, the large deviations approach preserves the thermodynamic constraints on the current statistics imposed by the fluctuation symmetry (Appendix \ref{Appendix : Fluctuation symmetry}). Here, we will consider another such constraint, the so-called Thermodynamic Uncertainty Relations (TUR) \cite{barato2015thermodynamic,horowitz2020thermodynamic, falasco2020unifying}, in the particular context of the CMOS inverter.
These relations express a fundamental trade-off between the precision (defined as the ratio between the mean and the standard deviation) of any current observable and the total entropy production. 
For example, for any empirical current $I \equiv N_\rho(t)/t$ in the long time limit, the bound states that
\begin{eqnarray}
\frac{\langle I \rangle^2}{\sigma^2_{I}}\leq \frac{\langle \Sigma_t \rangle}{2k_B}
\label{eqn: TUR}
\end{eqnarray}
where $\Sigma_t$ is the total entropy production of the system. For the CMOS inverter, $T \langle\Sigma_t\rangle =\langle I_{\rm{p/n}} \rangle\Delta V t$ corresponds to the associated average power dissipation. 
\par

In Figure~\ref{fig:av_var_current}-(c), we plot the precision per entropy production as a function of $V_{\rm{in}}$ 
for different powering voltages $V_{\rm{dd}}$. The mean and variance of the current were computed 
from the large deviations SCGF, and it satisfies the inequality imposed by TUR.  We also see that the bound imposed by the TUR is only approached close to equilibrium, i.e., for low values of $V_\text{dd}$. This can again be shown using the fluctuation symmetry, since the equality of the bound is achieved as we take $\Delta V \to 0 $, where both the variance and the average current become proportional to the conductivity $\Omega$ \cite{saryal2019thermodynamic}.
We also note that the precision per entropy production is 
maximized for $V_\text{in}=0$, when the timescales of both conduction channels match. Similar phenomenology has been observed in other systems where the kinetic asymmetry between channels leads to less precise currents \cite{ptaszynski2017nonrenewal,ubbelohde2013spin, bulka2000current,barato2015thermodynamic}. This general behaviour is also reproduced by our simple 2-state model in Appendix \ref{Appendix: 2 state}.

\subsection{Coarse-Grained model for the CMOS inverter}


Since electronic circuit engineering works in a top-down fashion, a complex circuit is designed starting with functional modules, implemented using elementary components \cite{itzkovitz2005coarse}. Even for a simple functional modules like the NOT gate, computing the current statistics with multiple transistors can be a challenging task. The difficulty arises as a result of the correlations between the jumps in different transistors. Here, we will discuss the conditions under which one can consider a coarse-grained model of the CMOS inverter, ignoring the details of individual transistors, while still correctly describing the current fluctuations.
 \par
Under such a coarse-graining procedure, we would model the charge transfer through the inverter as a single bi-Poissonian process. The current through the inverter is then defined using the net charge flow through it at any time interval $t$ as, $I(t)=\Delta q (t)/t = (q_e/t)[N_+(t)-N_-(t)]$, where $N_{\pm}$ are the individual Poisson processes with corresponding coarse-grained rates $\lambda^{\pm}_{\rm{cg}}$ (Fig.~\ref{fig:coarse-graining}(a)). In contrast to the individual transistors, the rates here would be obtained using the average current, $\langle I \rangle (\Delta V) = \langle \Delta q (t)/t \rangle= q_e [\lambda^+_{\rm{cg}}(\Delta V)-\lambda^-_{\rm{cg}}(\Delta V)]$, where $\Delta V =V_{\rm{dd}}-V_{\rm{ss}}$. For thermodynamic consistency, we will also enforce the local detailed balance condition $\log[\lambda^+_{\rm{cg}}(\Delta V)/\lambda^-_{\rm{cg}}(\Delta V)]= \Delta V/V_{\rm{T}}$. The advantage of using such an approximate modeling is that all higher-order cumulants can be obtained just from the measurement of the average current $\langle I \rangle (\Delta V)$. For example, the variance of the coarse-grained inverter can then be computed as $\sigma^2(I) = \frac{q_e^2}{t}[\lambda^+_{\rm{cg}}(\Delta V)+\lambda^-_{\rm{cg}}(\Delta V)] = \frac{q_e}{t}\langle I \rangle \coth (\Delta V/2V_{\rm{T}}). $\par
In Figure~\ref{fig:coarse-graining}, we plot the scaled variance of the current for varying input voltage, $V_{\rm{in}}$, from the coarse-grained estimate and compare it with the large deviations result. For $|V_{\rm{in}}/V_{\rm{T}}|>1$, the coarse-grained estimate matches well with the exact one.  In this limit, one of the transistors is heavily activated compared to the other. Hence, one can coarse-grain the inverter to just a single bi-Poissonian process, corresponding to the slower transistor. But such a procedure fails to capture the local minimum at $V_{\rm{in}}=0$, as the dependence of jumps between the different devices is neglected in the coarse-grained model. \par

\section{Conclusion \& Discussion}
In this article, we presented an application of the formal tools of large deviations theory to correctly compute the voltage and current fluctuations in low-power electronic circuits. As a case study, we considered a low-power CMOS inverter, or NOT gate, which is a basic primitive in electronic design. Starting from a single-electron description of the circuit dynamics, we showed how the macroscopic regime of operation can be approached by scaling the physical dimensions of the transistors. In this regime of operation, the thermal fluctuations satisfied a Large Deviations Principle, which we employed to describe the fluctuations around the deterministic dynamics. As shown, this approach goes beyond the limitations of the traditional diffusive approximations to capture the rarer fluctuations. We finally applied the framework to analyze the current fluctuations in the CMOS inverter, and compared our results with the bound of TUR. and also discussed the conditions under which a coarse-grained effective model for the NOT gate can capture these fluctuations.

The CMOS inverter studied here is among the simplest non-linear electronic circuits, with just a single fixed point in the deterministic limit. This implies that the dynamics remains ergodic even when the macroscopic limit is considered. Hence, the large size limit commutes with the long time limit for any dynamical observable. It will be interesting to study more complex electronic circuits, like memories with multiple fixed points \cite{freitas2021reliability}, or clocks or oscillators with limit cycles \cite{abidi2006phase}, where the commutativity of these limits may be broken \cite{jack2020ergodicity,vroylandt2020efficiency}. 

Our work shows how the tools of large deviations theory and stochastic thermodynamics can be applied to a family of electronic systems of high technological relevance. In that way, we open up a new playground to explore fundamental questions of non-equilibrium statistical physics. At the same time, our methods could contribute to the search and design of new computing schemes operating in low-power regimes \cite{han2013approximate,camsari2017stochastic, freitas2021stochastic}, where the effects of thermal fluctuations cannot be neglected.

\section{Acknowledgments}
A.G. thanks Jan Meibohm and Krzysztof Ptaszynski for the discussions and helpful comments. This research was supported by the project FNR CORE program, project NTEC (C19/MS/13664907) and INTER/FNRS/20/15074473 funded by F.R.S.-FNRS (Belgium) and FNR (Luxembourg), and by the FQXi foundation project FQXi-IAF19-05.

\balance
\bibliographystyle{unsrt}
\bibliography{main}

\appendix
\onecolumngrid

\section{Solution to the Fokker-Planck Approximation}
\label{Appendix : FP solution}
In this section, we will compute the exact solution to the Fokker-Planck equation, along with its large deviation rate function.
As shown in Sec.3.1, in the limit of $v_e\to0$, one can truncate the above Kramer-Moyal expansion to obtain the following Fokker Planck equation,
\begin{eqnarray}
     \frac{\partial }{\partial t}P(v,t)=-\frac{\partial}{\partial v}\left[F(v)P(v,t)-\frac{1}{2}\frac{\partial}{\partial v}(D(v)P(v,t))\right]+o(v_e^2)
     \label{apeqn:FP eqn}
\end{eqnarray}
where we define the drift term as, $F(v)=\sum_\rho\Delta_\rho( \omega_\rho(v,0)+v_e\partial_{v_e}\omega_\rho(v,0))=\mu(v)+v_e\sum_\rho \Delta_\rho\partial_{v_e}\omega_\rho(v,0)$ and the diffusion term $D(v)=v_e\sum_\rho \omega_\rho(v,0)$. The steady state density can be obtained by setting $\partial_t P(v,t)=0$. Assuming the density decays to 0 as $v\to\infty$,  the solution is given by
\begin{eqnarray}
     P_{\rm{st}}(v)\propto \exp{\left[-\Phi(v)\right]}\int_{-\infty}^v dv' \frac{\exp\left[\Phi(v')\right]}{D(v')}
     \label{apeqn: FP_solution}
\end{eqnarray}
In the case of CMOS inverter with $V_{\rm{dd}}=-V_{\rm{ss}}\; \&\; n=1$, we obtain $\Phi(v)$ as follows,
\begin{eqnarray}
      \Phi(v)&=&\int\left[\frac{2F(v')-\partial_v D(v')}{D(v')}\right]dv' \nonumber\\
     &=&  \frac{2V_{\rm{T}}}{v_e}\Bigg\{2\left[e^{2V_{\rm{in}}/V_{\rm{T}}}-1\right]\frac{\tan^{-1}\left[\frac{1+e^{2V_{\rm{in}}/V_{\rm{T}}}+2e^{(v-V_{\rm{dd}})/V_{\rm{T}}}}{\sqrt{4e^{2(V_{\rm{in}}-V_{\rm{dd}})/V_{\rm{T}}}-\left(1+e^{2V_{\rm{in}}/V_{\rm{T}}}\right)^2 }} \right]  } {\sqrt{4e^{2(V_{\rm{in}}-V_{\rm{dd}})/V_{\rm{T}}}-\left(1+e^{2V_{\rm{in}}/V_{\rm{T}}}\right)^2 }}\nonumber\\
     &+&\ln\left(e^{v/V_{\rm{T}}}+e^{2V_{\rm{dd}}/V_{\rm{T}}}+e^{(2V_{\rm{in}}-v)/V_{\rm{T}}}+e^{(2V_{\rm{in}}+V_{\rm{dd}})/V_{\rm{T}}}\right)\Bigg\}.
 \label{apeqn: exact_FP}
 \end{eqnarray}
As discussed in the main text, one can also use the large deviation ansatz, $P(v,t)\asymp \exp(-v_e^{-1}\phi_{\rm{FP}}(v,t))$, to obtain the rate function corresponding to Fokker Planck equation by solving 
\begin{eqnarray}
 d_t \phi_{\rm{FP}}(v,t) = \mu(v)\partial_v \phi_{\rm{FP}}(v)+d(v)[\partial_v \phi_{\rm{FP}}(v)]^2,
\end{eqnarray}
where we have defined $d(v)=D(v)/(2v_e)$. For the CMOS inverter, the stationary rate function is given by,
\begin{eqnarray}
 \phi_{\rm{FP}}(v)=\int\left[-\frac{\mu(v')}{d(v')}\right]dv'=v_e\Phi(v)
\end{eqnarray}
This means that the contribution to the steady state from the integral in Eq.~\eqref{apeqn: FP_solution} is sub-exponential in $v_e$.

\section{Path Integral Approach}
\label{app: path_integral}
In this section, we will briefly review the Martin-Siggia-Rose (MSR) path integral construction for the Markov jump processes. Then we will re-derive the large deviation results for the state observables (quasipotential approach) of the CMOS inverter in the macroscopic limit using the path integral methods. We will also develop a systematic way of getting the usual Gaussian approximation for both state and current observables through the path integral and discuss their limitations. We will be mainly following the approach developed in \cite{tomassothesis}, to set up the path integral tools.\par
Here, a trajectory is defined as a time-ordered sequence of charge on the output conductor, $[q_t]=\{q(\tau): 0\le \tau \le t\}$, such that $q(0)=q_0$.  Dividing the trajectory of length t into M equal slices of width $t_{i+1}-t_i=t/M$, one can obtain the probability of a trajectory $[q_t]$, as follows
\begin{eqnarray}
     P[q_t] &=& \prod_{i=0}^{M-1} \left\langle\delta\left(q(t_{i+1})-q(t_i)-q_e\sum_\rho \Delta_\rho \mathcal{N}_\rho(t_i)\right)\right\rangle_{\mathcal{N}}P(q_0,0)\nonumber\\
     &=& \prod_{i=0}^{M-1} \left\langle \int_{-\infty}^{\infty}\frac{dp(t_{i+1})}{2\pi}e^{-ip(t_{i+1})\left(q(t_{i+1})-q(t_i)-q_e\sum_\rho \Delta_\rho \mathcal{N}_\rho(t_i)\right)}\right\rangle_{\mathcal{N}}P(q_0,0)
\end{eqnarray}     
where $\mathcal{N}_\rho(t)$ is a Poisson noise with the associated rate $\lambda_\rho(t)$. Now taking the limit of $M\to\infty$ and rotating the variables $p(t_{i})\to i p(t_{i})$, we get the following functional path probability
\begin{eqnarray}
    P[q_t] &=& \int \mathcal{D}p \;e^{\mathcal{A}_t[q_t,p_t]}P(q_0,0) =: \int\mathcal{D}p \;P[q_t, p_t] 
    \label{appeq : path prob}
\end{eqnarray}
where we have defined the "action" functional $\mathcal{A}_t[q_t,p_t]$ as
\begin{eqnarray}
 \mathcal{A}_t[q_t,p_t]
    \equiv\int_0^t\left[ -p.\dot{q}+\sum_\rho(e^{p(\tau).\Delta_\rho q_e}-1)\lambda_\rho(q(\tau))\right]d\tau \nonumber,
\end{eqnarray}
Hence for each trajectory $[q_t]$, we can associate a path of the conjugate field $[p(t)]=\{p(\tau): 0\le \tau \le t\}$ accounting for the fluctuations needed to create that trajectory. One notes that the "kinetic" term, $\int_0^t-p.\dot{q}$ of $\mathcal{A}_t[q_t,p_t]$ accounts for the overlap between the states, $q$ and $p$. Identifying the usual Lagrangian structure to the action, one can also introduce the Hamiltonian functional, $H[q,p]$, which is the characteristic functional of Poisson distribution
\begin{eqnarray}
    H[q,p]&:=& \sum_\rho(e^{p(\tau).\Delta_\rho q_e}-1)\lambda_\rho(q(\tau)).\nonumber
\end{eqnarray}
The Hamiltonian functional has a similar structure to the Hamiltonian operator defined from the master equation. It is also the characteristic functional of Poisson distribution. 

\subsection{Quasipotential approach \& Gaussian approximation}
\label{app: LD_quasi}
Even though the path probability has this Hamiltonian structure, the exact solution of the path integral (Eq.~\eqref{appeq : path prob}) is usually not feasible analytically. Hence, we will discuss how analytical results can be obtained by taking the macroscopic limit, where one does a saddle-point approximation. Here, we will show that these are equivalent to the large deviation results obtained in Section.~\ref{sec: quasi}. Since the action sits on an exponential, the validity of using the saddle point approximation corresponds to the existence of the large deviation principle.\par
In the macroscopic limit $v_e \rightarrow 0$, it is natural to assume the charge in conductors also scales with size, and hence the voltage $v$ in the conductor becomes the natural variable. As discussed before, we define the rescaled Poisson rates $\omega_\rho(v,v_e)=v_e\lambda_\rho(q)$, such that $\lim_{v_e\to0}\omega_\rho(v,v_e)\equiv \omega_\rho(v)$ exists. The path probability for $v$ with rescaled rates $\lambda_\rho(q)=v_e^{-1}\omega_\rho(v,v_e)$ is given by
\begin{eqnarray}
    P[v_t]&=&\int\mathcal{D}p \;e^{v_e^{-1}\int_0^t\left[ -p.\dot{v}+\sum_\rho(e^{p(\tau).\Delta_\rho }-1)\omega_\rho(v(\tau),v_e)\right]}P(v_0,0)\nonumber\\
    &=&:\int\mathcal{D}p \; e^{v_e^{-1}[a_t[v,p]-\phi(v_0,0)]}
    \label{appeqn: pathprob_large}
\end{eqnarray}
where we have rescaled $v_e p\rightarrow p$, and without loss of generality assume a similar exponential scaling for initial distribution, $P(v_0,0)=e^{-v_e^{-1}\phi(v_0,0)}$. Extremizing the action $\frac{\delta a_t[v,p]}{\delta v}=0=\frac{\delta a_t[v,p]}{\delta p}$, one obtains the typical trajectories, following Hamilton's equations of motion in the extended space $(v,p)$:
\begin{eqnarray}
     \dot{v}(\tau) &=&\sum_\rho\Delta_\rho e^{\Delta_\rho p(\tau)}\omega_\rho(v(\tau))\label{apeqn: eom_v}\\
    \dot{p}(\tau) &=& -\sum_\rho\left( e^{\Delta_\rho p(\tau)}-1\right)\frac{\partial}{\partial v}\omega_\rho(v(\tau))
    \label{apeqn: eom_p}
\end{eqnarray}
The choice of boundary conditions is guided by the choice of observables.

In most of the applications, one is interested only in the stationary behaviour of the system. In the stationary state, the leading trajectories are those which nullify the Hamiltonian, i.e., the ones that pass extremely close to the fixed point. In the macroscopic limit, such leading trajectories become more typical and hence the equivalent condition to obtain the quasipotential is
\begin{eqnarray}
 H= \sum_\rho(e^{p.\Delta_\rho}-1)\omega_\rho(v)=0 
 \label{eqn:quasi_hamiltonian}
\end{eqnarray}  
whose solution provides the auxiliary variable $p=p(v)$, a function of v. Since the Hamiltonian $H[p,v]=0$, the resultant action is just $a_t[v,p]=-\int_0^t d\tau \;p\dot{v}=\int dv' p(v')$. One can also interpret these trajectories as the one which takes infinite time to go to the fixed point of dynamics, i.e. $v_{\text{det}}$ to final voltage $v$. Hence, the stationary distribution $P_{\rm{st}}(v)$ is given as
\begin{eqnarray}
     \lim_{t\to\infty}\langle\delta(v(0)-v_{\text{det}})\delta(v(t)-v)\rangle &\sim& e^{-v_e^{-1}\int_{v_{\text{det}}}^v p(v')dv'}\nonumber\\
     &=:& e^{-v_e^{-1}[\phi(v)-\phi(v_{\rm{det}})]}.
     \label{appeqn: Quasi_path}
\end{eqnarray}
where $\phi(v)$ is the quasipotential. In the case of CMOS inverter, the quasipotential is obtained by solving,
\begin{eqnarray}
    0 = (e^{p(\tau)}-1)[\omega_-^\text{n}(v) + \omega_+^\text{p}(v)] +(e^{-p(\tau)}-1)[\omega_+^\text{n}(v) + \omega_-^\text{p}(v)]
\end{eqnarray}
from which we get the auxiliary variable,  $p(v)=\partial_v \phi(v)= \ln\left[\frac{\omega_-^\text{n}(v) + \omega_+^\text{p}(v)}{\omega_-^\text{p}(v) + \omega_+^\text{v}(v)}\right]$. Solving the above equation, we get back the quasipotential obtained by assuming the large deviation ansatz in the master equation (Eq.~\eqref{eqn: quasipotential}).\par
Now we will derive the Fokker-Planck equation, obtained in Sec.~\ref{sec: quasi}, using the path integral approach. In the limit of $v_e\to  0$, the action may be approximated by truncating it to $o(v_e^{2})$ to obtain the Gaussian approximation of the Poisson noise. The rescaled action in Eq.~\eqref{appeqn: pathprob_large} is then given as
\begin{eqnarray}
     a_t[v,p]=\int_0^t d\tau  \;v_e\Bigg[p\left(\sum_\rho \Delta_\rho [\omega_\rho(v)+v_e\partial_{v_e}\omega_\rho(v)|_{v_e=0}]-\dot{v}\right)
     +\frac{v_e}{2}\sum p^2\left(\sum_\rho \omega_\rho(v)\right)+o(v_e^{2})\Bigg],
     \label{eqn:action_gaussian}
\end{eqnarray}
where we put back the rescaled conjugate field $p\to v_e p$, as done in Eq.~\eqref{appeqn: pathprob_large}. Now taking the Gaussian integral over p, one obtains the Onsager-Machlup action corresponding to the Langevin dynamics,
\begin{eqnarray}
     \dot{v} = F(v)+\sqrt{D(v)}\eta,
     \label{appeqn: langevin}
\end{eqnarray}
where  the drift term is $F(v)=\sum_\rho \Delta_\rho [\omega_\rho(v,0)+v_e\partial_{v_e}\omega_\rho(v,0)]$, the diffusion coefficient is $D(v):=v_e\sum_\rho\omega_\rho(v,0)$ and $\eta$ is the Gaussian white noise with zero mean and unit variance. Hence, we obtain back the Langevin dynamics corresponding to the Fokker-Planck approximation discussed before (Sec.~\ref{sec: quasi}).\par
One can also note that the deterministic dynamics is easily obtained by neglecting even the terms of $o(v_e)$ in the action, $a_t[v,p]$ of Eq.~\eqref{eqn:action_gaussian}, given as
\begin{eqnarray}
     \dot{v}= \mu(v) = \sum_\rho \Delta_\rho \omega_\rho(v,0),\qquad v(0)=V_0
\end{eqnarray}
where we have assumed $P(v,0)\sim\delta(v-v_0)$. All the information of the fluctuations are discarded while taking the $p=0$ limit. The fixed point in the corresponding dynamics is also the deterministic steady solution, calculated using Kirchhoff's law (Eq.~\eqref{eqn: deterministic}). \par

\subsection{Large deviation method for Current Statistics \& Gaussian approximations}
\label{app: LD_current}
In this section, we derive the different diffusive approximations by expanding the biased action (Eq.~\eqref{appen: MGF_path}) in $v_e$ and applying the saddle point approximation. We also compare the resultant SCGF from the different diffusive approximations with the full Hamiltonian (Eq.~\eqref{eqn: SCGF_Hamiltonian}).
As shown in Sec.~\ref{sec: current_macro}, the MGF of the integrated current $N^\rho_t$, in the path integral representation is given as,
\begin{eqnarray}
    \mathcal{Z}_\rho(\xi,t) &=& \left\langle\exp(\xi N_\rho)\right\rangle\nonumber\\
    &=&\int\mathcal{D}v\mathcal{D}p e^{v_e^{-1}\int_0^td\tau[-p.\dot{v}+H_\xi(v,p)]}P(v_0,0)
    \label{appen: MGF_path}
\end{eqnarray}
where one can now define a biased action $a^\xi_t[v,p] \equiv \int_0^td\tau[-p.\dot{v} +H_\xi(v,p)]$. Here, we have the biased Hamiltonian, defined as
\begin{eqnarray}
    H_\xi(v,p)\equiv\sum_{\rho'}\left(e^{p.\Delta_{\rho'} + s(\rho')\xi}-1\right)\omega_{\rho'}(v,v_e)
    \label{appeqn: biasedH}
\end{eqnarray}
where $s(\rho)$ is the sign of $\rho$ whose current is being computed, else $0$ for $\rho\ne\rho'$, and $\xi$ is the counting field associated with the device current. \par
\begin{figure}[h!]
\centering
\includegraphics[width=\textwidth]{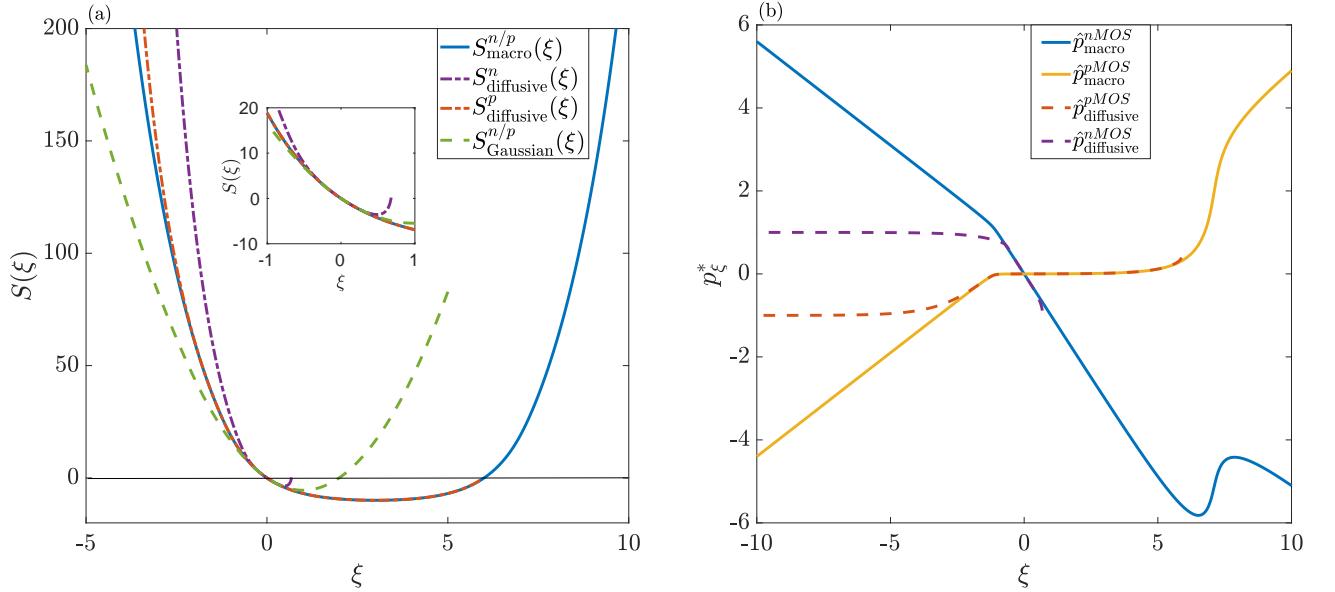}
\caption{(a) The scaled cumulant generating function $S(\xi)$ for the CMOS inverter, obtained using different approximations. Here, 'macro' is the one computed using the full Hamiltonian, 'diffusive' corresponds to the one expanding the action only in $p$ up to $2^{\rm{nd}}$ order, and the 'Gaussian' to the one where the action is expanded in both $p$ and $\xi$ up to $2^{\rm{nd}}$ order.  The curves for diffusive approximations are cut-off at certain values of $\xi$, as the fixed points are lost in such approximations. (b) The fixed points $p^*$ , for the nMOS and pMOS transistors, corresponding to the full Hamiltonian one in Eq.~\eqref{apeqn: eom_v}-\eqref{apeqn: eom_p} is compared with the diffusive approximation one in Eq.~\eqref{apeqn: diffusive_vxi}-\eqref{apeqn: diffusive_pxi}. From the values of the fixed points in the macroscopic limit, one can identify the performance of the diffusive approximation for the different devices. Parameters : $V_{\text{dd}}/V_{\rm{T}}=3, V_{\text{in}}/V_{\text{dd}}=0.5, v_e/V_{\rm{T}}=0.1$. }
\label{fig:SCGF}
\end{figure}

The diffusive approximation for the current through a device $\rho$ can be obtained by expanding the action in Eq.~\eqref{appen: MGF_path} up to $o(v_e^{2})$. The biased action for the integrated current through the device $\rho$ in the limit of $v_e\to0$ is then given as,
\begin{flalign}
a^\xi_t[v,p] \!=\!\int_0^t \!d\tau\! \Bigg[\!\sum_\rho\!\left(\!e^{s(\rho)\xi}-1\!\right)\!\omega_\rho(v)
\!+\!p\,v_e\Big(\sum_\rho\Delta_\rho e^{s(\rho)\xi}[\omega_\rho(v)\!+\!v_e^2\partial_{v_e}\omega_\rho(v,0)]\!-\!\dot{v}\Big)+\frac{(p\, v_e)^2}{2}\sum_\rho\omega_\rho(v)e^{s(\rho)\xi} + o(v_e^2)\!\Bigg]\nonumber
\end{flalign}
Notice that this is just equivalent to expanding the biased Hamiltonian $H_\xi(v,p)$ in $p$ up to $o(p^2)$, as shown in the main text (Eq.~\eqref{eqn: biasedH_gauss1}). Comparing it with the previous section, one can note that we are effectively doing the full current statistics for the Langevin dynamics of Eq.~\eqref{appeqn: langevin}. After putting $p\, v_e\to p$, we can again extremize the action to get the following equations of motion,
\begin{flalign}
\dot{v}&=\sum_\rho\left[(\Delta_\rho+p)e^{s(\rho)\xi}\right]\omega_\rho(v,0)\label{apeqn: diffusive_vxi}\\
\dot{p}&=-\sum_\rho\left[(1+p\Delta_\rho+p^2/2)e^{s(\rho)\xi}-1\right]\frac{\partial}{\partial v}\omega_\rho(v,0).
\label{apeqn: diffusive_pxi}
\end{flalign}
As seen in the main text, the above diffusive approximation fails to capture the symmetry of the biased Hamiltonians (Eq.~\eqref{symmetry_pn}), thereby becoming nonphysical.
A consistent way of doing the diffusive approximation, respecting the symmetries of the system, is to expand the action (Eq.~\ref{appen: MGF_path}) in both $p$ and $\xi$ up to second order. Such an approximation leads to the Hamiltonian of Eq.~\eqref{eqn: biasedH_gauss2}, and hence captures only the Gaussian fluctuations in the current.

In Figure~\ref{fig:SCGF}-(a), we compare the SCGF obtained from the different approximations for both devices of the CMOS inverter. For both the macroscopic approximations with the complete Hamiltonian (Eq.~\eqref{eqn: biased_H}) and the proper diffusive approximation (Eq.~\eqref{eqn: biasedH_gauss2}), the symmetry is satisfied and hence $S_n(\xi)=S_p(\xi)$. Whereas, one can see that the symmetry of the biased Hamiltonian (Eq.~\eqref{symmetry_pn}) is broken in the diffusive approximation of Eq.~\eqref{eqn: biasedH_gauss1}. It is interesting to note that this diffusive approximation for the pMOS current statistics captures the macroscopic one for a larger range of counting field $\xi,$ whereas it performs poorly for the nMOS transistor. This could be justified by the fact that the approximation of $p\to0$ fails in the case of the nMOS transistor, even for small values of the counting field $\xi$, compared to the pMOS transistor (Figure~\ref{fig:SCGF}-(b)). As with the state observable, the diffusive approximation is again thermodynamically inconsistent. Both of them fail to capture the fluctuation symmetry of the SCGF, i.e., $S(\xi)=S(-\xi-\beta q_e \Delta V)$. This can be seen as none of the diffusive approximations captures the zero of $S_\xi$, corresponding to the thermodynamic affinity $\beta q_e \Delta V$.

\section{Minimum of Rate function is the Deterministic Limit}
\label{Appendix: Minimum rate function}

When we define the large deviation rate function based on the large size limit, i.e. $v_e\to0$, we are essentially taking the deterministic limit. Hence, the minimum of the rate function of the state (quasi-potential), and the rate function of any integrated currents must be at the deterministic value. This implies that the most probable value even in the stochastic regime at this limit is the deterministic case. But since the rate function captures non-Gaussian fluctuations, implies that the average value of the state/currents could be different from the deterministic value at finite size.
\par
In the case of state distribution, it is easy to check this from Eq.~\eqref{appeqn: Quasi_path}, as we showed that the quasipotential has the form,
\begin{eqnarray}
 \phi(v) = \int_{v_{\text{det}}}^v p(v')dv'.
\end{eqnarray}
From this it is clear that the minimum of the rate function/quasipotential is when $\phi'(v)=p(v)=0$, which is the deterministic case. 
\par
In the case of current statistics, let us first consider the scaled cumulant generating function, given by
\begin{eqnarray}
  S(\xi)=-v_e^{-1}H_\xi[v^*_\xi,p^*_\xi]= -v_e^{-1}\sum_\rho\left(e^{p^*_\xi.\Delta_\rho+s(\rho)\xi}-1\right)\omega_\rho(v^*_\xi)
\end{eqnarray}
where $p^*_\xi, v^*_\xi$ are optimal fixed points of biased dynamics of Eq.~\eqref{eqn: SCGF_Hamiltonian}. From Gartner-Ellis theorem, one can identify the large deviation rate function for current, $I(J)$, by
\begin{eqnarray}
  I(J)= \max_\xi\left\{\xi J -S(\xi)\right\}
  \label{Legendre_current}
\end{eqnarray}
where the maximization is equivalent to  solving the equation, $J=\partial S/\partial\xi|_{\xi^*}$, to obtain the following equation
\begin{eqnarray}
 J =   -v_e^{-1}\sum_\rho\left( \frac{\partial p^*_\xi}{\partial \xi}.\Delta_\rho+s(\rho)\xi\right) e^{p^*_\xi.\Delta_\rho+s(\rho)\xi}\omega_\rho(v^*_\xi)+\left(e^{p^*_\xi.\Delta_\rho+s(\rho)\xi}-1\right)\frac{\partial \omega_\rho(v^*_\xi)}{\partial \xi},
 \label{Current_CF}
\end{eqnarray}
Now, the minimum of the rate function is given when $\frac{\partial I}{\partial J}=0$, implies,
\begin{eqnarray}
 \frac{\partial \xi^*(J)}{ \partial J } J+ \xi^*(J) = \frac{\partial S}{\partial \xi}|_{\xi^*}\frac{\partial\xi^*}{\partial J}\\
 \xi^*(J) = \left( \frac{\partial S}{\partial\xi}\Big|_{\xi^*} -J \right)\frac{\partial\xi^*}{\partial J}
\end{eqnarray}
Hence, the value of the counting field at the minimum of rate function at $\Tilde{J}$, is $\xi^*(\Tilde{J})=0$. Now putting everything back into Eq.(\eqref{Current_CF}), we get
\begin{eqnarray}
 \Tilde{J} =v_e^{-1}\sum_\rho\left(\Delta_\rho \frac{\partial p^*}{\partial\xi}\Big|_{\xi^*=0}+s(\rho)\right)\omega_\rho(v_{\rm{det}})
\end{eqnarray}
where the first term vanishes as $v_{\rm{det}}$ is the fixed point of unbiased dynamics, i.e. $\dot{v}=\sum_\rho\Delta_\rho\omega_\rho(v_{\rm{det}})=0$. Hence, we get that the minimum of rate function is at $\Tilde{J}=v_e^{-1}\sum_\rho s(\rho)\omega_\rho(v_{\rm{det}})$, which is the deterministic solution of Kirchhoff's law (Eq.~\ref{eqn: detkitchoff}). This is due to the fact that the rescaled rate does not have the contribution to the rate due to the charging effect at the dominant order of $v_e$.

\section{Proof of Fluctuation symmetry in the SCGF}
\label{Appendix : Fluctuation symmetry}
Here we will derive the fluctuation symmetry using the symmetry in the biased generator and path integral techniques. As shown in Sec.~\ref{sec: Spectral}, the biased generator can be decomposed as
\begin{eqnarray}
    \hat{\mathcal{L}}_\xi &=&\hat{\gamma}(t)-\hat{\Gamma}_\xi(t) \\ 
    \hat{\gamma}(t) &=&\sum_q\ket{q}\bra{q}\left(-\sum_\rho\lambda_\rho(q,t)\right) \hspace{0.15cm}\rm{and} \hspace{0.15cm} \hat{\Gamma}_\xi(t)=\sum_q\sum_\rho\ket{q}\bra{q-q_e\Delta_\rho }\lambda_\rho(q-q_e\Delta_\rho)e^{\xi_{|\rho|}s(\rho)}\nonumber
\end{eqnarray}
Now one can show that there exists a symmetry for $\hat{P}_{\text{eq}}^{-1}\hat{\Gamma}_\xi\hat{P}_{\text{eq}}=\hat{\Gamma}^T_{\Bar{\xi}}$, where $\hat{P}_{\text{eq}}$ is a diagonal operator with equilibrium distribution, $P_{\text{eq}}(q)=Z^{-1}e^{-\beta\psi(q)}$,  with $\psi(q)$ being the generalized energy function as defined in [],
\begin{eqnarray}
   \hat{P}_{\text{eq}}^{-1}\hat{\Gamma}_\xi\hat{P}_{\text{eq}}= \sum_q\sum_\rho\ket{q}\bra{q-q_e\Delta_\rho }\lambda_\rho(q-q_e\Delta_\rho)e^{\xi_{|\rho|}s(\rho)}e^{-\beta\left[\psi(q-q_e\Delta_\rho)-\psi(q)\right]}
\end{eqnarray}
Now using the energy balance for the jumps in a trajectory, $\psi(q)-\psi(q-\Delta_\rho q_e)=\delta Q_\rho(q-\Delta_\rho q_e)+s(\rho)q_e\Delta V$ and then using the local detailed balance condition,
\begin{eqnarray}
   \frac{\lambda_\rho(q-\Delta_\rho q_e)}{\lambda_{-\rho}(q)}=e^{-\beta\delta Q_\rho(q-\Delta q_e)}
\end{eqnarray}
we get,
\begin{eqnarray}
  \hat{P}_{\text{eq}}^{-1}\hat{\Gamma}_\xi\hat{P}_{\text{eq}}&=& =\sum_q\sum_\rho\ket{q}\bra{q-q_e\Delta_\rho }\lambda_{-\rho}(q)e^{s(\rho)\left[\xi_{|\rho|}+\beta q_e \Delta V\right]}\nonumber\\
  &=&\sum_q\sum_{-\rho}\ket{q}\bra{q+q_e\Delta_\rho }\lambda_{\rho}(q)e^{s(\rho)\left[-\xi_{|\rho|}-\beta q_e \Delta V\right]}\nonumber\\
  &=&  \hat{\Gamma}^T_{-\xi_{|\rho|}-\beta q_e \Delta V}\nonumber
\end{eqnarray}
This symmetry implies that the moment generating function, $Z(\xi,t)=\bra{1}e^{-t \hat{L}_{\xi}}\ket{P_{\text{eq}}}= \rm{Tr}[e^{t\hat{L}_{\xi}}\hat{P}_{\text{eq}}]$ has the following symmetry $Z(\xi,t)=Z(-\xi-\beta q_e\Delta V,t)$ (Lebowitz-Spohn symmetry).\par
\begin{eqnarray}
P(N_t)/P(-N_t)=e^{\beta q_e \Delta V N_t}
\end{eqnarray}

Using the path integral technique, we will show that the symmetry in the biased Hamiltonian corresponds to the symmetry in the scaled cumulant generating function, $S(\xi)$. The biased Hamiltonian for the statistics of the integrated current through a device $\rho$, in the limit of $v_e\to0$ limit, is 
\begin{eqnarray}
H_\xi(v,p)&=&\sum_\rho \omega_\rho(v)\left[e^{\Delta_\rho p-\xi_{|\rho|}s(\rho)}-1\right]
= \sum_{\rho}\omega_{-\rho}(v)\left[e^{-\Delta_\rho p+\xi_{|\rho|}s(\rho)}-1\right]\\
&=&\sum_{\rho}\omega_{\rho}(v)\left[e^{-\Delta_\rho p+\xi_{|\rho|}s(\rho)+\beta \delta Q_\rho(v)}-1\right]
\end{eqnarray}
In the large size limit, the first law for a transition $\rho$ at the trajectory level can be simplified, $\delta Q_\rho(v)=\Delta_\rho \frac{\partial}{\partial v}\Tilde{\Psi}_\rho(v)+s(\rho)q_e \Delta V$, where we have used $\Psi(q)=v_e^{-1}\Tilde{\Psi}(v)+o(v_e^{-1})$. 
\begin{eqnarray}
H_\xi(v,p)=\sum_{\rho}\omega_{\rho}(v)\left[e^{-\Delta_\rho p+\xi_{|\rho|}s(\rho)+\beta( \Delta_\rho \frac{\partial}{\partial v}\Tilde{\Psi}_\rho(v)+s(\rho)q_e \Delta V)}-1\right]=H_{-\xi-\beta q_e \Delta V}\left(v,-p+\beta\frac{\partial}{\partial v}\Tilde{\Psi}_\rho(v)\right).\nonumber
\end{eqnarray}
Hence, the Hamiltonian, $H_\xi(v,p)$ is invariant under transformation
\begin{eqnarray}
p\rightarrow-p+\beta\frac{\partial}{\partial v}\Tilde{\Psi}_\rho(v),\qquad \xi \rightarrow-\xi-\beta q_e\Delta V
\end{eqnarray}
Since the rates are time independent, using the time reversal transformation, $\tau\to t-\tau$, along with the above transformation, one can obtain the following symmetry for the cumulant generating function, $\log Z(\xi,t)$ in the large time limit,
\begin{flalign}
&\log Z(\xi,t)=\log \int \mathcal{D}v \int \mathcal{D}p \exp\left[v_e^{-1}\int_0^t d\tau(-p.\dot{v}+H_\xi(v,p))\right]P_{\rm st}(v_0)\nonumber\\
 &= \log\int \mathcal{D}v \int \mathcal{D}p \exp\left[v_e^{-1}\int_0^t d\tau\left(-p.\dot{v}-\dot{v}\beta\frac{\partial}{\partial v}\Tilde{\Psi}(v)+ H_{-\xi-\beta q_e \Delta V}\left(v,-p+\beta\frac{\partial}{\partial v}\Tilde{\Psi}_\rho(v)\right)\right)\right]P_{\rm st}(v_{\rm{T}})\nonumber\\
&=\log Z(-\xi-\beta q_e \Delta V,t)+o(t)
\end{flalign}
where the subleading terms are due to the boundary terms due to stationary distribution and energy function. Hence, we obtain the back fluctuation symmetry for the scaled cumulant generating function of current, $S(\xi)=S(-\xi-\beta q_e \Delta V)$.

\section{2 State Model with 2 Channels}
\label{Appendix: 2 state}

\begin{figure}[h!]
\centering
\includegraphics[width=0.8\textwidth]{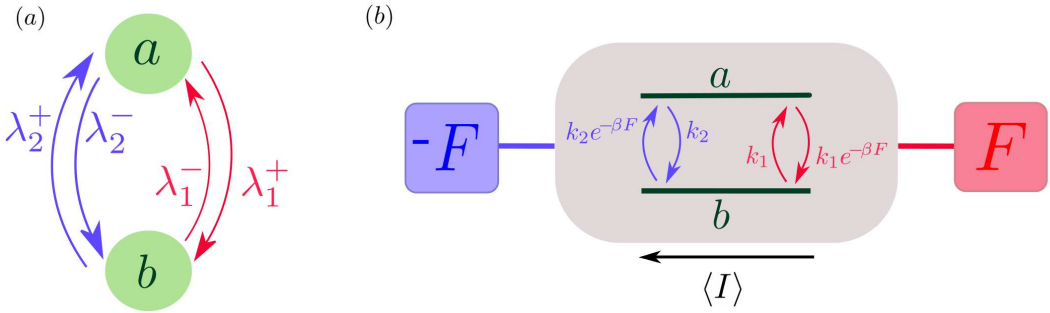}
\caption{(a): 2 state Markov jump process with 2 channels (with different kinetic activities). (b): Each channel, 1 (red) and  2 (blue), has jumps due to a bath with an associated thermodynamic affinity, $-\beta F$ and $\beta F$ respectively. Hence, there is an average current flow in the steady state of the system.}
\label{fig: 2state}
\end{figure}

Consider a system  with 2 degenerate states, $\boldsymbol{x}\equiv\{a,b\}$, where the dynamics is governed by a Markov jump process. Here, the jumps can occur through 2 different channels $\{1 \;\&\; 2\}$, with associated thermodynamic affinities $\beta F$ and $-\beta F$ respectively (Figure~\ref{fig: 2state}(b)). This is motivated by the CMOS inverter where the pMOS and nMOS are in contact with reservoirs at potential $V_{\rm{dd}}$ and $-V_{\rm{dd}}$. Each channel is modelled as a bidirectional Poisson process, with associated rates $\lambda_{1/2}^\pm$ proportional to their kinetic activity, $k_{1/2}$.  The master equation describing the evolution of the probability vector $P(x,t)$ for the above model (Figure~\ref{fig: 2state}(a)) is
\begin{eqnarray}
d_t P(\boldsymbol{x},t) = \hat{L}, P(\boldsymbol{x},t)
\end{eqnarray}
where the generator $\hat{L}$ is given as
\begin{eqnarray}
\hat{L} = 
\begin{bmatrix}
-(\lambda_1^+ + \lambda_2^-) & (\lambda_1^- + \lambda_2^+) \\
(\lambda_1^+ + \lambda_2^-)  & -(\lambda_1^- + \lambda_2^+)
\end{bmatrix}.
\end{eqnarray}
Using the local detailed balance condition and choosing the kinetic constants for the channels as $k_1 \;\&\; k_2$ respectively, we can define the rates (without loss of generality) as follows:
\begin{eqnarray}
\lambda_1^+ &=& k_1 ,\qquad\qquad \;\lambda_2^+ = k_2, \\
\lambda_1^- &=& k_1 e^{-\beta F}; \qquad \lambda_2^- = k_2 e^{-\beta F} ;
\end{eqnarray}

The stationary state for the above problem is given as,
\begin{eqnarray}
P_{\rm{st}}(a)=\frac{(\lambda_1^- + \lambda_2^+)}{\lambda_1^+ + \lambda_2^-+\lambda_1^- + \lambda_2^+},\qquad P_{\rm{st}}(b)=\frac{(\lambda_1^+ + \lambda_2^-)}{\lambda_1^+ + \lambda_2^-+\lambda_1^- + \lambda_2^+}
\end{eqnarray}

We are interested in understanding the long time current statistics through a channel when we change the timescale (activity) of each channel. Hence, we will compute the SCGF for different relative kinetic activity of the channels, i.e. $k_r=k_2/k_1$.  We also constrained the rates, such that the timescale associated with the flux from the bath with higher affinity to the lower one, $\tau_0=1/\sqrt{k_1k_2}$, be fixed. In the CMOS inverter, this constraint is the result of the fact that the input voltage ($V_{\rm{in}}$) exponentially activates \& suppresses the conduction through the transistors, i.e. $\lambda_{p/n} \propto e^{\pm V_{\rm{in}}}$. The current statistics through the channel 2 can be obtained using the titled generator (rescaled by $\tau_0$), $\hat{L}_\xi$
\begin{eqnarray}
\hat{L}_{\xi} = \frac{1}{\sqrt{k_1k_2}}
\begin{bmatrix}
-(\lambda_1^+ + \lambda_2^-) & (\lambda_1^-  + \lambda_2^+e^{\xi}) \\
(\lambda_1^+  + \lambda_2^-e^{-\xi})  & -(\lambda_1^- + \lambda_2^+)
\end{bmatrix}
\end{eqnarray}
In the long time limit, the current statistics is given by the eigenvalue with the largest real part of $\hat{L}_{\xi}$, which is the SCGF. The exact expression for the SCGF $S_1(\xi)$ is 
\begin{eqnarray}
S_2(\xi)= \frac{1}{2\sqrt{k_r}}\left[-(1+e^{-\beta F})(1+k_r)+\sqrt{2e^{-\beta F}\left[(1+k_r)^2+(1-k_r)^2\cosh{(\beta F)}+4k_r\cosh{(\beta F+\xi)}\right]}\right]
\end{eqnarray}
In the such a limit, the current statistics though both the channels are equal, and hence we can define $S(\xi)\equiv S_1(\xi)=S_2(\xi) $. Notice that the fluctuation symmetry $S(\xi)=S(-\xi-2\beta F)$ is also satisfied. The mean and the scaled variance of the current, in the units of $\tau_0$, are then given as
\begin{eqnarray}
\langle I \rangle &=& \frac{ \sqrt{k_r} (1-e^{-\beta F})}{(1+kr)} \label{apeqn : 2state
_av}\\
\sigma^2(I)t &=& \frac{2e^{-2\beta F}\sqrt{k_r}(2k_r+(1+k_r^2)\cosh{(\beta F)})}{(1+e^{\beta F})(1+k_r)^3}
\label{apeqn : 2state
_var}
\end{eqnarray}

\begin{figure*}[h!]
\centering
\includegraphics[trim=0 220 0 10, clip,width=\textwidth]{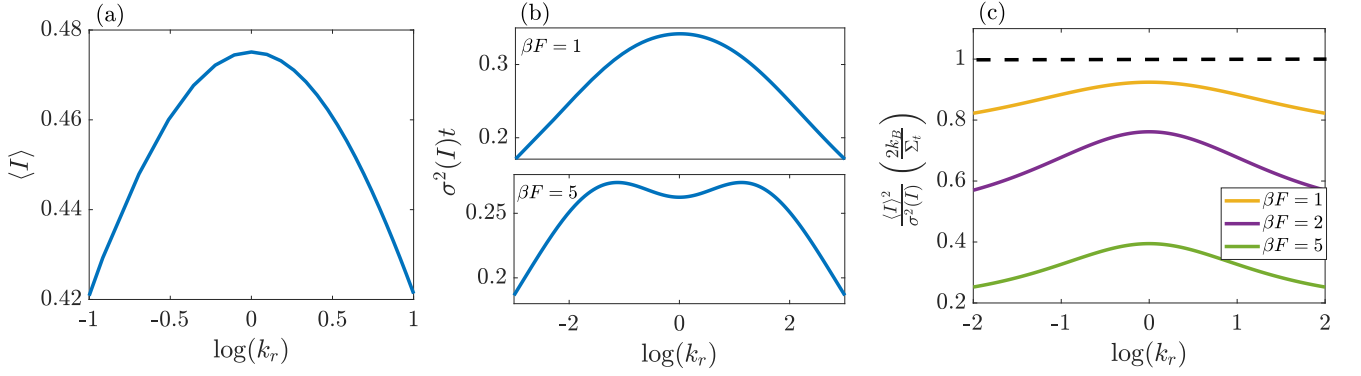}
\caption{Fig: (a) The mean current through channel 1 in the long time limit. (b) The scaled variance of current through channel 1 in the long time limit. There is an interesting phenomenology that the variance is not maximum when timescales match. Parameter: $\beta F =5$. (c) The precision of the current scaled by entropy production is maximum when the timescales match. The TUR also becomes tighter, as we decrease the affinity $F$.}
\label{fig:av_var_2state}
\end{figure*} 

We now use perturbation methods to characterize the behavior of the current statistics in the different limits of $\log(k_r)$. Since we are interested in the long time statistics, we need to perturb the titled generator to compute the largest eigenvalue. For example, in the case of $k_r=1+\epsilon$, where $\epsilon\to0$, one can do a perturbation analysis of the tilted generator,
\begin{eqnarray}
\hat{L}_{\xi} = (\hat{L}_1+\hat{L}_{2\xi})+\frac{\epsilon}{2}(\hat{L}_{2\xi}-\hat{L}_1),
\end{eqnarray}
where $\hat{L}_1=L_1/k_1$, $\hat{L}_{2\xi}=L_{2\xi}/k_2$ are the scaled generators corresponding to each channel. Using the traditional perturbation tools, one can obtain the following series expansion for the SCGF $S(\xi)$,
\begin{eqnarray}
S(\xi) = \left[-(1+e^{-\beta F})+e^{-\xi/2}(1+e^{-\beta F-\xi})\right]+\frac{(1+e^{-2\beta F})-e^{\xi}(1+e^{-2\beta F-2\xi})}{2e^{\xi/2}(1+e^{-\beta F-\xi})}\epsilon^2+\mathcal{O}(\epsilon^3).
\end{eqnarray}
We find that the first order contribution to SCGF is 0, and hence there will be a local extremum for all cumulants of the current at $k_r=1$. The scaled variance expanded up to $o(\epsilon^2)$ is
\begin{eqnarray}
\sigma^2(I)t = \frac{1+e^{-\beta F}}{4}+\frac{\cosh{\beta F}-3}{16(1+e^{\beta F})}(k_r-1)^2+\mathcal{O}(k_r-1)^3.
\end{eqnarray}
From the above equation, one can notice that the scaled variance goes from being local maximum ($\cosh{\beta F}<3$) to a local minimum ($\cosh{\beta F}>3$) at $k_r=1$. 


Now let us look at the limit of $|\log(k_r)|\to\infty$. In such a limit, the current statistics of the system will be controlled by the slower channel. Using this information, the tilted generator in the case of $k_r\to0$ can be perturbed as follows,
\begin{eqnarray}
\hat{L}_{\xi}=\frac{1}{\sqrt{k_r}}\left(\hat{L}_1+ k_r \hat{L}_{2\xi}\right)
\end{eqnarray}
where we have assigned the counting field to the slower channel 2. Computing the series expansion for the $S(\xi)$ as before, one can obtain the following perturbation series for the variance as, $k_r\to0$
\begin{eqnarray}    
\sigma^2(I)t = \frac{2\cosh{\beta F}}{(1+e^{\beta F})}\sqrt{k_r}+\frac{2(2-3\cosh{\beta F})}{(1+e^{
\beta F})}k_r^{3/2}+\mathcal{O}(k_r^{5/2})
\end{eqnarray}
where the first term is the contribution to the current from the slower channel 2 assuming the system has equilibrated due to the faster channel 1. This implies that the zeroth order term vanishes as the largest eigenvalue of $\hat{L}_1$ is 0, corresponding to its equilibrium state. Here, we are effectively doing a timescale separation, such that the population is controlled by the faster channel and the current statistics is controlled by the slower channel. Similarly, one can also obtain the perturbation series for $k_r\to\infty$ limit, where we need to tilt the generator of channel 1 (slower one).

In Figure~\ref{fig:av_var_2state}, we plot the mean, scaled variance, and precision/cost of the current through a channel with the relative kinetic activity of the channels $k_r$. In Figure~\ref{fig:av_var_2state}-(a), we see that the mean current is maximum when the timescales of the channels match. From Eq.~\eqref{apeqn : 2state
_av}, we can see that $\langle I \rangle \propto [\sqrt{k_r}/(1+k_r)]$, which is the equal to the conductivity $\Omega(1/\tau_0)=d \langle I \rangle/d(\beta F)|_{\beta F\to 0}$ of the system.
The decaying behavior of $\langle I \rangle$ with $k_r$ is due to the fact that the conductivity of the full system is controlled by the conductivity (timescale) of the slower channel, which goes as $e^{-|\log(k_r)|/2}$ for $|\log(k_r)|\gg1$. In Figure~\ref{fig:av_var_2state}-(b), we plot the scaled variance in the long time limit for two cases: near equilibrium case ($\beta F=1$) and far from equilibrium ($\beta F=5$). Near equilibrium, the behaviour is similar to that of the conductivity $\Omega$, which determines the equilibrium variance (as shown in Sec.~\ref{sec: CMOS_current}). Far from equilibrium $\beta F\gg 1$, we see that the variance is maximized at finite $\log(k_r)$, as seen also in the CMOS inverter.  
The combination of both these cumulants can be captured by the precision by cost, which is known to be bounded from above through the thermodynamic uncertainty relation (TUR) (Eqn.\eqref{eqn: TUR}). We find that the precision per cost of the current is maximum when both channels are equally active (Figure~\ref{fig:av_var_2state}-(c)). For this system, the cost corresponds to the total entropy production $\Sigma_t= \langle I \rangle (2\beta F) t $, which is just the average current $\langle I \rangle$ times the affinity of the system ($2\beta F$). One can also see that the performance of the bound deteriorates as we go far from equilibrium, as the equality for TUR bound is achieved as we converge to equilibrium. We see that the key features of the CMOS inverter are all captured by this simple 2-state model, and hence they are a result of just the competition between the timescales of the 2 transistors, which is controlled by $V_{\rm{in}}$.
\par

\end{document}